\documentclass[prx,superscriptaddress,aps,amsmath,amssymb,floatfix,reprint,raggedbottom]{revtex4-2}

\usepackage{graphicx}
\graphicspath{{Figures/}}
\usepackage{pgfplots}
\pgfplotsset{compat=1.15}

\usepackage{balance}
\usepackage{xr-hyper}
\usepackage{enumitem}
\usepackage{wrapfig}

\usepackage{amssymb}
\usepackage{amsmath}
\usepackage{times}

\usepackage{dcolumn}
\usepackage{xcolor}

\usepackage{lipsum}
\usepackage{soul}
\usepackage{braket}
\usepackage{amsthm}
\usetikzlibrary{shapes.gates.logic.US, positioning}
\usepackage{amsfonts}
\usepackage{multirow}
\usepackage{tcolorbox}
\usepackage{bbm}
\usepackage{blkarray} 

\usepackage[utf8]{inputenc}
\usepackage{hyperref}

\usepackage[linesnumbered,ruled,vlined]{algorithm2e}

\DeclareUnicodeCharacter{2009}{\,}

\makeatletter
\newcommand*{\addFileDependency}[1]{
  \typeout{(#1)}
  \@addtofilelist{#1}
  \IfFileExists{#1}{}{\typeout{No file #1.}}
}
\makeatother

\makeatletter 
\renewcommand{\fnum@figure}{\textbf{Fig.~\thefigure}}
\makeatother

\makeatletter
\def\bbordermatrix#1{\begingroup \m@th
  \@tempdima 4.75\p@
  \setbox\z@\vbox{
    \def\cr{\crcr\noalign{\kern2\p@\global\let\cr\endline}}
    \ialign{$##$\hfil\kern2\p@\kern\@tempdima&\thinspace\hfil$##$\hfil
      &&\quad\hfil$##$\hfil\crcr
      \omit\strut\hfil\crcr\noalign{\kern-\baselineskip}
      #1\crcr\omit\strut\cr}}
  \setbox\tw@\vbox{\unvcopy\z@\global\setbox\@ne\lastbox}
  \setbox\tw@\hbox{\unhbox\@ne\unskip\global\setbox\@ne\lastbox}
  \setbox\tw@\hbox{$\kern\wd\@ne\kern-\@tempdima\left[\kern-\wd\@ne
    \global\setbox\@ne\vbox{\box\@ne\kern2\p@}
    \vcenter{\kern-\ht\@ne\unvbox\z@\kern-\baselineskip}\,\right]$}
  \null\;\vbox{\kern\ht\@ne\box\tw@}\endgroup}
\makeatother

\emergencystretch=\maxdimen
\usepackage[compact]{titlesec}
\titlespacing{\section}{0pt}{*3}{*2}
\titlespacing{\subsection}{0pt}{*2}{*2}
\titlespacing{\subsubsection}{0pt}{*2}{*2}

\titleformat{\section}{\filcenter\normalfont\small \bfseries}{\thesection.}{1em}{}   

\newcommand{\beginsupplement}{

        \setcounter{table}{0}

        \renewcommand{\thetable}{S\arabic{table}}
        
        \setcounter{algocf}{0}
        \renewcommand{\thealgocf}{S\arabic{algocf}}
        
        \setcounter{figure}{0}
        \renewcommand{\thefigure}{S\arabic{figure}}
        \setcounter{equation}{0}
        \renewcommand{\theequation}{S.\arabic{equation}}
     }

\usepackage{booktabs}

\begin{document}

\title{Generalized Probabilistic Approximate Optimization Algorithm}
\par

\author{Abdelrahman S.  Abdelrahman}
\affiliation{Department of Electrical and Computer Engineering, University of California, Santa Barbara, Santa Barbara, CA, 93106, USA}
\author{Shuvro Chowdhury}
\affiliation{Department of Electrical and Computer Engineering, University of California, Santa Barbara, Santa Barbara, CA, 93106, USA}

\author{Flaviano Morone}
\affiliation{Center for Quantum Phenomena, Department of Physics, New York University, New York, New York 10003 USA}
\author{Kerem Y. Camsari}
\affiliation{Department of Electrical and Computer Engineering, University of California, Santa Barbara, Santa Barbara, CA, 93106, USA}

\begin{abstract}
We introduce the generalized Probabilistic Approximate Optimization Algorithm (PAOA), a classical variational Monte Carlo framework that extends and formalizes the recently introduced PAOA, enabling parameterized and fast sampling on present-day Ising machines and probabilistic computers. PAOA operates by iteratively modifying the couplings of a network of binary stochastic units, guided by cost evaluations from independent samples. We establish a direct correspondence between derivative-free updates and the gradient of the full Markov flow over the  exponentially large state space, showing that PAOA admits a principled variational formulation. Simulated annealing emerges as a limiting case under constrained parameterizations, and we implement this regime on an FPGA-based probabilistic computer with on-chip annealing to solve large 3D spin-glass problems. Benchmarking PAOA against QAOA on the canonical 26-spin Sherrington--Kirkpatrick model with matched parameters reveals {superior performance} for PAOA. We show that PAOA naturally extends simulated annealing by optimizing multiple temperature profiles, leading to improved performance over SA on heavy-tailed problems such as SK--Lévy.
\end{abstract}
\maketitle

\section{Introduction}
\label{sec:Intro}

Monte Carlo algorithms remain a central tool for exploring complex energy landscapes, especially in the context of combinatorial optimization and statistical physics. Classical methods such as simulated annealing (SA) have been widely applied across these domains, but their reliance on slowly equilibrating processes limits their performance on rugged energy landscapes~\cite{wang2015ising, shen2024fem, munozarias2024clifford, ma2024mpvan, barzegar2024annealing, fan2023dirac}. New approaches are needed to construct non-equilibrium strategies that retain algorithmic simplicity while improving solution quality. 

Inspired by the quantum approximate optimization algorithm (QAOA) ~\cite{QAOA_max_cut_farhi, Herrman2022,vijendran2024expressive, MISHA_Lukin_QAOA_2020, mueller2024limitations, montanaro2024speedup, zhou2020qaoa, boulebnane2025equivalence, gulbahar2023maximum, finvzgar2024designing, cheng2024quantum}, Weitz et al. \cite{Combes_2025} proposed a classical variational protocol based on the direct parameterization of low-dimensional Markov transition matrices. Their work introduced the term Probabilistic Approximate Optimization Algorithm (PAOA), raising the possibility of classical, variational analogs to QAOA within probabilistic architectures.

Building on this foundational work, we formalize and  generalize PAOA. We move beyond the original proposal's edge-local matrices to derive a global \(2^N \times 2^N\) Markov-flow formulation applicable to any \(k\)-local Ising Hamiltonian of size $N$. This framework unifies a wide spectrum of variational ansätze, from global and local schedules to fully-parameterized couplings, and allows them to be stacked to an arbitrary depth \(p\). Crucially, we connect this variational theory to practice by building on the {p-computing framework}~\cite{camsari2019pbits}, where networks of binary stochastic units (p-bits) sample from Boltzmann-like distributions through asynchronous dynamics. This model has a demonstrated record of success in optimization and inference tasks~\cite{borders2019integer, aadit2022massively, nikhar2024all, chowdhury2023machine} and provides a natural substrate for our work. By implementing these dynamics on an FPGA, we demonstrate for the first time a scalable, hardware-based path for executing the PAOA.

We show that PAOA admits a broad class of parameterizations, including global, local, and edge-specific annealing schedules. Within this framework, SA emerges as a limiting case under constrained schedules. We implement this regime on an FPGA-based p-computer to demonstrate large-scale, high-throughput sampling. In contrast to standard SA, PAOA's flexible parameterization enables the discovery of non-equilibrium heuristics that exploit structural features of the problem. One such example is demonstrated in heavy-tailed Sherrington--Kirkpatrick (SK) models, where PAOA assigns higher effective temperatures to strongly coupled nodes and lower temperatures to weakly connected ones, enabling annealing profiles with better performance over vanilla SA. This adaptive scheduling capability provides a powerful framework for the automated discovery of novel annealing heuristics.

We also benchmark PAOA against QAOA on the SK model using matched parameter counts and observe superior performance.  This may not be surprising, as the practical performance advantages of QAOA over classical and probabilistic alternatives remain uncertain. A clear quantum advantage is expected to emerge in problems where interference plays an explicit role, such as in the synthetic examples constructed by Montanaro et al.~\cite{montanaro2024speedup}.  Our work builds on the promising results of Weitz et al.~\cite{Combes_2025} on the max-cut problem by expanding PAOA into a scalable, hardware-compatible, and general-purpose optimization algorithm.
\begin{figure*}[t]
  \centering
  \includegraphics[width=\linewidth]{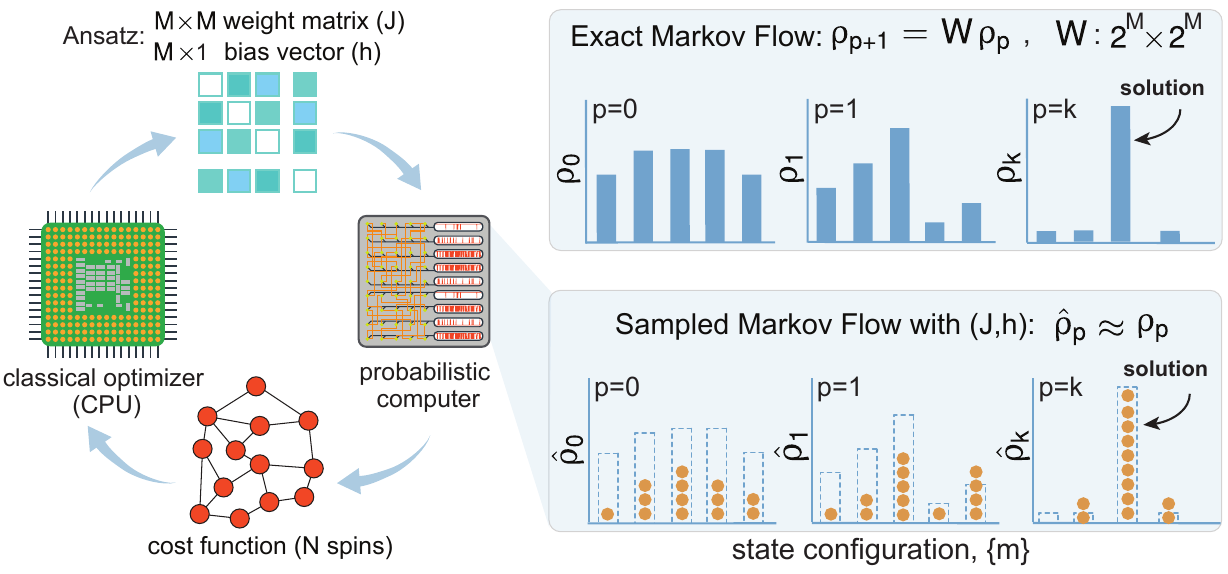}
  \caption{\textbf{Overview of PAOA}. A hybrid classical--probabilistic architecture iteratively updates a weight matrix $J$ and bias vector $h$ using feedback from a probabilistic computer. The p-computer samples from a distribution defined by $(J,h)$, approximating the exact Markov flow. The resulting samples $\hat{\boldsymbol{\rho}}_p$ are used to evaluate a cost function, which the classical optimizer minimizes by adjusting the variational parameters. Here, $\boldsymbol{\rho}$ denotes the exact probability distribution over spin configurations, $\hat{\boldsymbol{\rho}}$ the sampled approximation, $p$ the layer number, $M$ the number of total spins represented in the ansatz (including possible hidden variables), and $\{m\}$ a specific spin configuration (state). The cost function is typically the energy of a spin glass mapped to an optimization problem (e.g., Eq.~\eqref{eq3}) but can also be a likelihood function if PAOA is learning from data (e.g., Eq.~\eqref{cost_function}). Importantly, the ansatz size $M$ can exceed the original problem size $N$ since hidden variables may be introduced to increase the representation. In this work, we use $M=N$ in all experiments unless noted. At convergence ($p=k$), the distribution concentrates around low-energy solutions.}
  \label{fig:1}
\end{figure*}

The remainder of this paper is organized as follows: we first present the theoretical formulation of PAOA, including its parameterization, Markov structure, and sampling-based approximation with p-bits. In order to illustrate the correspondence between analytical gradients and derivative-free optimization, we study a simple majority gate problem. We then demonstrate the recovery of SA as a special case and implement this regime on hardware, followed by a description of the FPGA architecture used to accelerate sampling. Using matched parameter counts, we benchmark PAOA against QAOA on the SK model. Finally, we study a heavy-tailed SK variant to show how PAOA can discover multiple annealing schedules that outperform standard single-schedule simulated annealing.

\section{Probabilistic Approximate Optimization Algorithm Overview}
\label{sec:PAOA_theory}

The power of the Probabilistic Approximate Optimization Algorithm (PAOA) comes from a conceptual departure from traditional optimization methods, such as simulated annealing (SA). While SA explores a single, {fixed energy landscape} defined by a model Hamiltonian, PAOA treats the landscape itself as a {variational object} to be optimized. The goal is to dynamically reshape this landscape to make the ground state of the original problem easier to find.

This creates a two-level optimization loop, as illustrated in Fig.~\ref{fig:1}. In an outer loop, a classical optimizer adjusts a set of variational parameters, $\boldsymbol{\theta}$. In an inner loop, a probabilistic computer samples states from the variational landscape currently parameterized by $\boldsymbol{\theta}$. In the most general case, $\boldsymbol{\theta}$ specifies the full set of couplings and fields $\{J^{(k)},h^{(k)}\}$ at each layer $k$, so that the landscape itself is directly reconfigured by the optimizer. Simpler ansätze, such as global or node-wise schedules introduced in the Spectrum of variational ansätze subsection, can be viewed as constrained subsets of this general $J,h$ parameterization (for example, a global inverse temperature amounts to scaling all entries of $J$ by a single parameter). The sampler acts on $M$ spins in the ansatz, while the original problem has $N$ visible spins. In general $M\ge N$ because hidden spins may be introduced to enrich the representation, in direct analogy to neural quantum states (NQS) \cite{carleo2017solving}.  Unless otherwise stated, we set $M$=$N$ in the remainder of the paper, for simplicity, and leave the possibility of introducing hidden spins for better representation to future work. 

During training, samples generated under $\boldsymbol{\theta}$ are evaluated with a task-dependent cost function $\mathcal{L}(\boldsymbol{\theta})$. In problems with a known target set of states (e.g., majority logic gate, discussed in Section \ref{sec:Majority_vote}), $\mathcal{L}$ is typically chosen as a negative log-likelihood. In optimization problems such as 3D spin glasses, $\mathcal{L}$ is instead taken as the average energy over sampled configurations. In either case, feedback from the samples updates $\boldsymbol{\theta}$, decoupling the problem energy from the sampling dynamics and enabling PAOA to discover non-equilibrium pathways to solutions.

\subsection*{A Spectrum of Variational Ansätze}
The power and flexibility of PAOA lie in the choice of the variational parameters \(\boldsymbol{\theta}\) that define the search landscape. We refer to this parameterization as the variational ansatz. Let $N$ denote the total number of spins in the problem. To explore the tradeoffs between expressiveness and generalization, in this paper we study four representative ansätze, defined by the number of free parameters, \(\Gamma\):
\[
\Gamma \in \left\{1,\ 2,\ N,\ \frac{N(N-1)}{2} \right\}
\]
\begin{itemize}
    \item When $\Gamma$\,=\,$1$, the landscape is scaled by a single, learned inverse temperature \(\beta^{(k)}\) at each layer \(k\). This is the {global annealing schedule}, which is the closest analog to SA, with the crucial difference that the schedule is learned rather than predefined.
    \item When $\Gamma$\,=\,$2$, two independent schedules are assigned to distinct parts of the graph, typically divided based on node properties like weighted degree.
    \item When $\Gamma$\,=\,$N$, each node is assigned a {local schedule} \(\beta_i^{(k)}\), allowing the algorithm to anneal different parts of the problem at different rates.
    \item When $\Gamma$\,=\,$N(N-1)/2$, the entire symmetric coupling matrix \(J_{ij}^{(k)}\) is used as the variational ansatz, giving the algorithm maximum freedom to reshape the landscape at each layer.
\end{itemize}
Recent work in QAOA has shown that increasing the number of variational parameters per layer can improve convergence and reduce the required circuit depth~\cite{Herrman2022, vijendran2024expressive}. A similar principle applies to PAOA: more expressive schedules allow the system to reach useful non-equilibrium distributions in fewer layers (e.g., a single layer for a fully parameterized AND gate, as we show in Supplementary Figs.~\ref{fig:AND_gate} and \ref{fig:AND_gate_results}), but may also increase the risk of overfitting. The balance between expressiveness and generalization remains a key consideration when selecting an ansatz.

\subsection*{Implementation via Probabilistic Computers}
The practical implementation of the sampling inner loop is achieved using a probabilistic computing (p-computing) framework~\cite{camsari2019pbits}. This framework uses networks of binary stochastic units, or p-bits, that sample from Boltzmann-like distributions via asynchronous updates governed by Glauber dynamics. A single p-bit updates its state \(m_i\) according to
\begin{equation}
    m_i = \text{sgn}[\tanh(\beta I_i) - \text{rand}_{\text{u}}(-1, 1)]
    \label{eq1}
\end{equation}
where \(\text{rand}_{\text{u}}(-1, 1)\) is a random variable uniformly distributed in the range $[-1,1]$, \(\beta\) is the inverse temperature from the variational ansatz, and the local input \(I_i\) is
\begin{equation}
    I_i = \sum_j J_{ij} m_j + h_i
    \label{eq2}
\end{equation}
As long as connected p-bits are updated sequentially, in any random order,  this update rule generates samples from a Boltzmann distribution \(P(\{m\}) \propto \exp[-\beta E(\{m\})]\) over time, associated with the energy
\begin{equation}
    E(\{m\}) = -\sum_{i<j} J_{ij} m_i m_j - \sum_i h_i m_i
    \label{eq3}
\end{equation}

\subsection*{Formalism and connection to Markov chains}
While in practice PAOA relies on direct MCMC sampling, the process has a rigorous foundation in the theory of Markov chains. The evolution of the probability distribution \(\boldsymbol{\rho}\) over the state space can be described by a sequence of transition matrices. To formalize this, we consider a \(p\)-layer process where the distribution after \(k\) layers, \(\boldsymbol{\rho}_k\), is given by
\begin{equation}
    \boldsymbol{\rho}_k = W^{(k)}(\boldsymbol{\theta}^{(k)}) \cdots W^{(1)}(\boldsymbol{\theta}^{(1)}) \boldsymbol{\rho}_0
    \label{finite_Markov_chain}
\end{equation}
where \(\boldsymbol{\rho}_0 \in \mathbb{R}^{2^N}\) is the initial distribution and each \(W^{(k)}\) is a \(2^N \times 2^N\) transition matrix parameterized by the variational parameters \(\boldsymbol{\theta}^{(k)}\) for that layer. For a sequential update scheme, each \(W^{(k)}\) can be factorized as a product of single-site update matrices, \(W^{(k)} = w_N^{(k)} w_{N-1}^{(k)} \cdots w_1^{(k)}\). The exact construction of these matrices from the p-bit update rule is detailed in Supplementary Section 3.

This layered application of stochastic matrices to a probability vector \(\boldsymbol{\rho}\) is the direct classical analog of applying unitary operators \(U^{(k)}\) to a wavefunction \(\psi\) in QAOA. The crucial distinction, however, is that each \(W^{(k)}\) is a norm-one-preserving stochastic matrix that mixes non-negative probabilities, whereas each \(U^{(k)}\) is a norm-two-preserving unitary matrix that rotates complex vectors. 
\begin{figure*}[t!]
    \centering
    \includegraphics[width=1\linewidth]{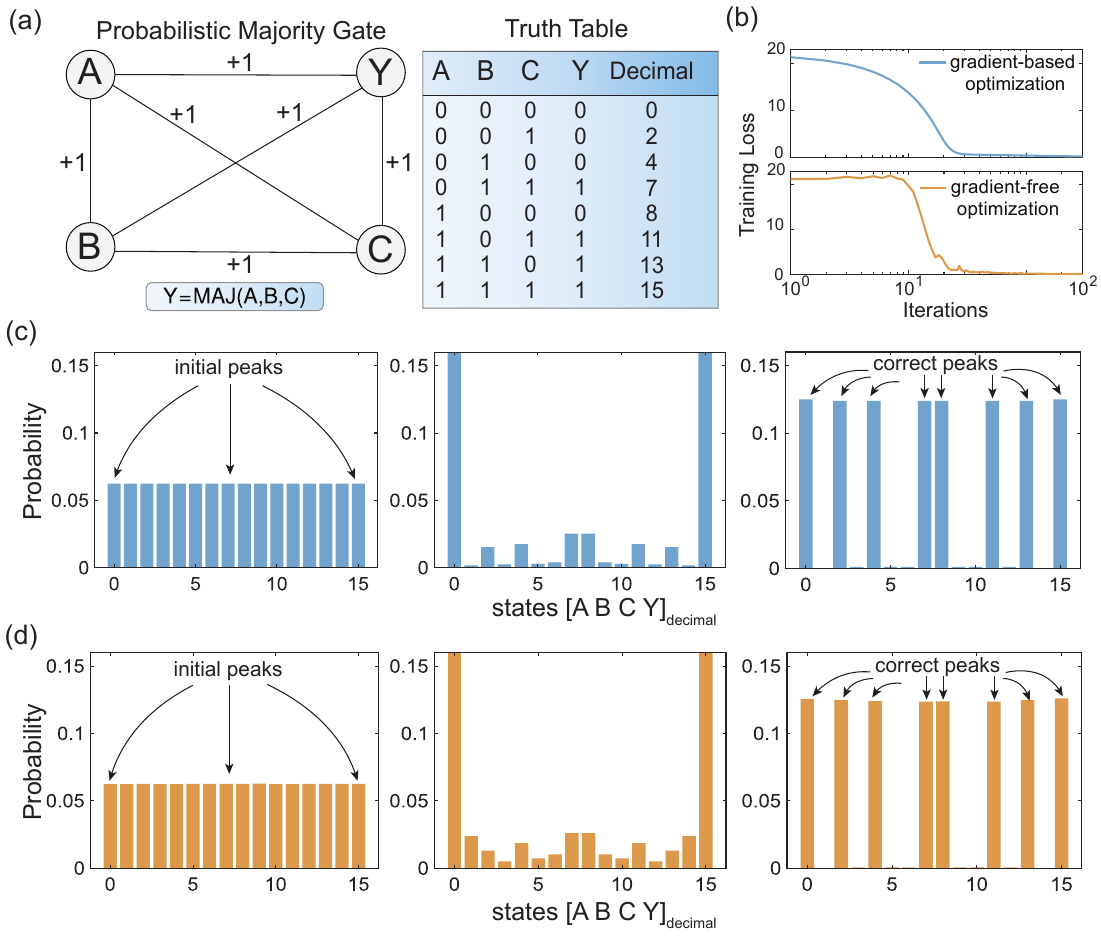}
    \caption{{\footnotesize \textbf{Majority gate benchmark: comparing analytical and sampling-based PAOA}.
    (a) Fully connected four-node network used to implement the majority gate $Y$\,=\,$\text{MAJ}(A,B,C)$, where $Y$\,=\,$A \lor B$ if $C$\,=\,$1$ and $Y$\,=\, $A \land B$ if $C$\,=\,$0$. The table lists the eight valid input-output combinations, labeled by their decimal encoding.
    (b) Training loss during optimization, comparing exact gradients (blue) to gradient-free optimization using COBYLA (orange) with $10^7$ MCMC samples. Both methods converge to the same minimum.
    (c) Time evolution of the exact distribution over two layers using analytical Markov matrices. Initial uniform distribution ($p$\,=\,$0$) is transformed into a peaked distribution ($p$\,=\,$2$) concentrated on the correct truth table entries.
    (d) Corresponding evolution using MCMC samples and COBYLA. The approximated distributions closely match the exact dynamics.}} 
    \label{fig:majority_gate}
\end{figure*}

The absence of complex phases in the classical evolution means that PAOA proceeds without the possibility of quantum interference. Consequently, for optimization problems where a constructive interference mechanism for QAOA is unclear, its advantage over classical alternatives is not guaranteed. Indeed, as we will demonstrate in Section \ref{SK_model}, our benchmarking shows that PAOA consistently reaches higher-approximation ratios on the Sherrington-Kirkpatrick model when compared to QAOA using an identical number of variational parameters.

The ultimate objective of PAOA is to find the states $\{m\}$ that minimize the problem's energy function (Eq.~\eqref{eq3}). This is achieved through the two-level optimization loop described earlier. The outer loop minimizes a cost function (different from the problem's energy function) over the variational parameters $\boldsymbol{\theta}$, thereby reshaping the sampling landscape to make the optimal states $\{m\}$ more probable and easier to find through MCMC sampling in the inner loop.

For a target set of states \(\mathcal{X}\), this cost function 
can be the negative log-likelihood:
\begin{equation}
    \mathcal{L}(\boldsymbol{\theta}) = -\sum_{\{m\} \in \mathcal{X}} \ln(\boldsymbol{\rho}_p(\{m\}; \boldsymbol{\theta}))
    \label{cost_function}
\end{equation}
In practice, as the exact computation of \(\boldsymbol{\rho}_p\) is intractable for large \(N\), it is replaced by an empirical distribution \(\hat{\boldsymbol{\rho}}_p\) estimated from \(N_E\) independent MCMC samples. Because the samples are generated from a shallow Markov process (small \(p\)), the system typically remains out of equilibrium. This non-equilibrium character distinguishes PAOA from classical annealing methods and may unlock new features such as initial condition dependence and faster convergence to the solution. However, this may also become a liability, for shallow circuits, as strong dependence on the initial distribution \(\boldsymbol{\rho}_0\) or overfitting may prevent the learned heuristics from generalizing. Therefore, the same \(\boldsymbol{\rho}_0\) used during training should also be used during inference.  {Deep} PAOA circuits do not exhibit any initial condition dependence especially with small initial $\beta$ that randomizes spins (see Section \ref{sec:discovery_SA}).

Finally, the PAOA framework is general in two key respects. First, the {cost function} being minimized is not restricted to the two-local form, such as the Ising energy of Eq.~\eqref{eq3}; it can be any computable function over the states \(\{m\}\), including higher-order \(k\)-local Hamiltonians or likelihood functions for machine learning tasks. Second, the {variational landscape} used for sampling, while implemented here with an Ising-like p-computer, is also not fundamentally limited. The PAOA approach is compatible with any model from which one can efficiently draw samples via MCMC, such as Potts models.

\section{Representative Example:  Majority Gate }
\label{sec:Majority_vote}

To illustrate the behavior of PAOA under both exact and approximate dynamics, we solve a small optimization problem involving a four-node majority gate. This problem serves as a tractable testbed for understanding the role of the classical optimizer. 

The majority gate is defined over four binary variables \( [A, B, C, Y] \), where the output \( Y \) is given by
\begin{equation}
    Y = \text{MAJ}(A,B,C) =  \begin{cases}
        A \lor B, & \text{if } C = 1 \\
        A \land B, & \text{if } C = 0
    \end{cases}
    \label{MAJ_gate}
\end{equation}
 
 The eight correct input-output combinations, shown in the truth table in Fig.~\ref{fig:majority_gate}a, define the target set of states, \(\mathcal{X}\). The goal of the optimization is to find variational parameters that cause the final distribution, \(\boldsymbol{\rho}_p\), to be concentrated on this set.

As a starting graph, we consider a fully connected Ising graph with $J_{ij}$\,=\,$+1$  and no biases. In this example, the variational parameters are node-specific schedules \( \beta_i(p) \), corresponding to $\Gamma$\,=\,$N$. To validate the role of the numerical optimizer, we compare two optimization strategies: a full Markov chain with gradient-based optimization and an MCMC-based approximation using the COBYLA algorithm~\cite{powell1994cobyla}.

For the gradient-based approach, we use the formulation laid out in the Formalism and connection to Markov chains subsection and optimize over two layers ($p$\,=\,$2$) with a uniform initial distribution  $\boldsymbol{\rho}_0$\,=\,${1}/{16}$. The cost is the negative log-likelihood over the eight correct states, and the variational parameters \( \beta_i(p) \) are updated using gradient descent with a fixed learning rate $\eta$\,=\,$0.004$. Optimization stops when either the gradient norm falls below \( 10^{-7} \) or a fixed iteration budget is reached.

In the MCMC-based approach, the same initial condition is used, but sampling is performed using \( 10^7 \) independent runs. COBYLA is used to update the parameters based on the empirical distribution \( \hat{\boldsymbol{\rho}} \). The optimization stops based on parameter convergence or a maximum number of function evaluations. Results are shown in Fig.~\ref{fig:majority_gate}b, where the two curves converge to nearly identical minima.

Fig.~\ref{fig:majority_gate}c and d show the final distributions under both methods. The agreement between the exact Markov flow and the sampled histogram confirms that derivative-free optimization closely tracks the true gradient.

\begin{figure}[htp!]
    \centering
    \begin{algorithm}[H]
    \DontPrintSemicolon
\SetKwInOut{Input}{Input}
\SetKwInOut{Output}{Output}
\caption{PAOA: global annealing schedule}
\label{alg:numerical beta-parameterized-PAOA}
\Input{number of nodes $N$, number of layers $p$, number of sweeps per layer $s/p$, number of experiments $N_E$, problem weight matrix $J$, initial variational parameters $\mathbf{\beta}$, variational parameters tolerance $\varepsilon_{\text{step}}$, maximum iterations $t_{\max}$}
\Output{trained annealing schedule $\beta_{opt}$}
\SetKwFunction{FsubMain}{p-computer(FPGA)}
\SetKwFunction{FMain}{PAOA-circuit}
\SetKwProg{Fn}{Function}{:}{}

\Fn{\FsubMain{$\beta$, $J$, $N$, $p$, $s/p$, $N_E$}}{
    \For{$i\gets 1 $ \KwTo $N_E$}{
    initialize all spins randomly\;
    \For{$j \gets 1$ \KwTo $p$}{
        $\beta \gets \beta(i)$\;
        \For {$k \gets 1$ \KwTo $s/p$}{
            \For{$l \gets 1$ \KwTo $N$}{
            solve equations \eqref{eq1} and \eqref{eq2}
        }
        }
        
    }
    store p-bit states in the BRAM\;
    }

    \Return{all stored p-bit states}
    }
    \Fn{\FMain{$\beta$, $J$, $N$, $p$, $s/p$, $N_E$}}{
        
        states $\gets$ \FsubMain{$\beta$, $J$, $N$, $p$, $s/p$, $N_E$}\;
        compute the energy using equation \eqref{eq3} for all states\;
        compute the average energy\;
        \Return{average energy}
        }
        
        \While{(step size $>\varepsilon_{\text{step}}$ and number of iterations $<t_{\max}$)}{
        average energy $\gets$ PAOA-circuit ($\beta$, $J$, $N$, $p$, $s/p$, $N_E$)\;
        minimize average energy and get a perturbation vector ($\mathbf{p}$) using a gradient-free optimizer\;
        \For{$i \gets 1$ \KwTo $p$}{
        $\beta(i)^{t+1} \gets \beta(i)^{t} + \mathbf{p}(i)$
    }
   $t\gets t+1$
        }    
    
    \Return{optimal variational parameters}

\end{algorithm}
\end{figure}

We also test the same procedure on a 5-node full-adder circuit using the fully-parameterized \( J_{ij}(p) \) ansatz. The resulting distributions (see Supplementary Fig.~\ref{fig:full_adder_sequential_update}) once again demonstrate tight correspondence between exact and approximate optimization at each layer. 

Although analytical optimization is feasible for small \( N \), the \( 2^N \times 2^N \) transition matrix becomes intractable for larger systems. The success of COBYLA in this setting confirms that PAOA can be implemented efficiently on general p-computers using standard MCMC.
\begin{figure*}[t!]
    \centering
    \includegraphics[width=1\linewidth]{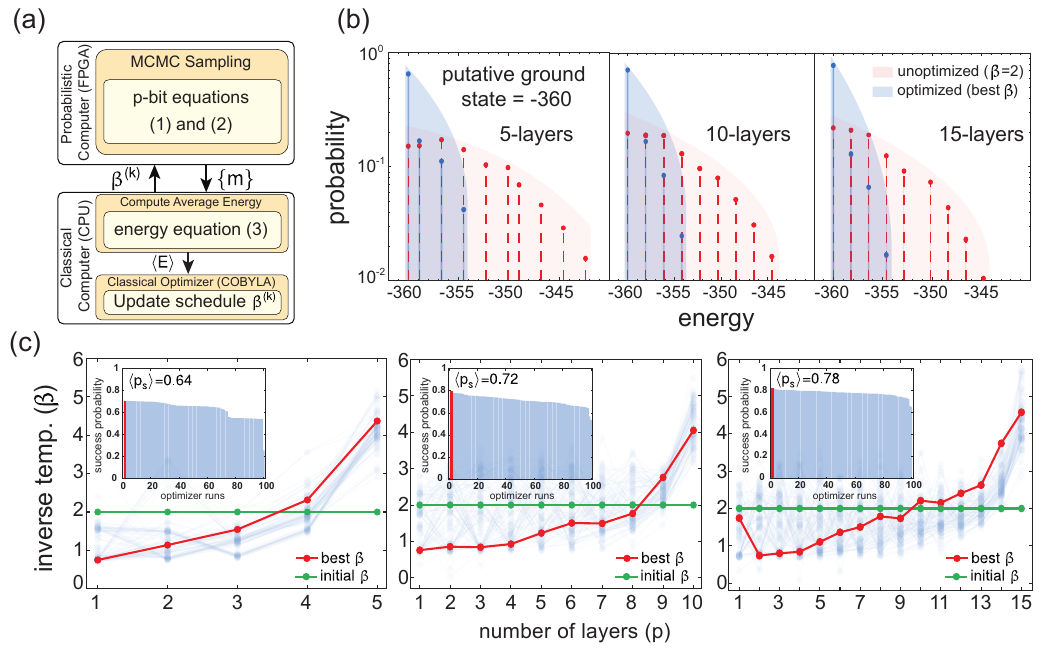}
    \vspace{-10pt}
    \caption{{\footnotesize \textbf{Discovering simulated annealing with PAOA on a 3D spin-glass problem}. (a) The hybrid architecture combines an FPGA-based p-computer for MCMC sampling, where $\{m\}$ represents the sampled state, with a classical CPU that optimizes the global annealing schedule (\(\beta^{(k)}\)) using the average energy ($\langle E \rangle$) computed across independent experiments/runs. (b) Energy histograms on a single $L^3$\,=\,$6^3$ instance before ($\beta$=$2$, red) and after (best $\beta$ schedule, blue) optimization. The optimized schedule shifts the distribution (shaded area) toward the putative ground state (energy\,=\,$-360$), increasing its discovery frequency out of $N_E$=$10^5$ independent runs. (c) Optimized schedules for \(p\)-layer architectures where \(p \in \{5, 10, 15\}\). Starting from a flat initial schedule ($\beta$\,=\,$2$, green), PAOA consistently discovers cooling schedules (best shown in red) that resemble SA. Faint curves show all 100 optimization runs. The insets display the sorted success probabilities, demonstrating that deeper architectures improve the average success probability ($\langle \mathrm{p}_s\rangle$) and reduce run-to-run variability.}}
    \label{fig:SA_discovery}
\end{figure*}

Interestingly, PAOA can explore parameter regimes not accessible to traditional Ising formulations. For example, optimal \( \beta \) values may be negative, corresponding to negative temperatures without a clear interpretation in the statistical physics-based context. Since equilibrium Boltzmann-sampling is not necessarily the starting point, alternative and more hardware-friendly activation functions could also be explored with PAOA, such as replacing the hyperbolic tangent in Eq.~\eqref{eq1} with the error function \(\mathrm{erf}\), or with saturating linear functions.

The shallow nature of the Markov chain used in PAOA introduces a dependency on the initial distribution $\boldsymbol{\rho}_0$. In contrast to standard Boltzmann training methods such as Contrastive Divergence~\cite{hinton2002training}, which seek to approximate the equilibrium distribution through extensive sampling, PAOA optimizes the evolution of the distribution under a fixed initialization. This shortcut avoids the computational burden of long mixing times but restricts the learned dynamics to a specific initialization at shallow depth. While this dependence can hinder generalization at shallow depth, as we demonstrate in Section \ref{sec:discovery_SA}, for deep PAOA circuits, initialization dependence is not an important concern.

\section{Discovering Simulated Annealing with PAOA}
\label{sec:discovery_SA}

An intriguing question is whether PAOA, when restricted to a global annealing schedule with one parameter per time step, can recover the well-known structure of simulated annealing (SA). In this section, we show that the answer is {yes}. Using a minimal parameterization with a single global inverse temperature, $\beta^{(k)}$, for each one of the $p$ total layers, PAOA is able to discover SA-like profiles that optimize the average energy on a large three-dimensional (3D) spin-glass instance.

\begin{table}[htp!]
    \centering
    \caption{Parameters used in Algorithm 1.}
        \vspace{1pc}
    \label{tab2:simulation_parameters}
    \renewcommand{\arraystretch}{1.2} 
    \setlength{\tabcolsep}{8pt} 
    \begin{tabular}{@{}ll@{}}
        \toprule
        Parameter & Value \\ 
        \midrule
        number of nodes ($N$) & 216  \\ 
        number of Monte Carlo Sweeps (MCS) /layer ($s/p$) & 720  \\ 
        number of layers ($p$) & \{5, 10, 15\}  \\ 
        initial variational parameters ($\mathbf{\beta}(p)$) & $2$  \\ 
        variational parameters tolerance ($\varepsilon_{\text{step}}$) & $10^{-4}$ \\ 
        maximum iterations ($t_{\max}$) & $5000$ \\ 
        number of independent experiments ($N_E$) & $10^5$ \\ 
        \bottomrule
    \end{tabular}
\end{table}

We apply PAOA to a 3D cubic lattice of size $L^3$\,=\,$6^3$, which yields a sparse bipartite interaction graph. To leverage this structure for massive parallelism on our FPGA-based p-computer, we use chromatic updates~\cite{aadit2022massively, nikhar2024all}. We employ the simplest $\Gamma$\,=\,$1$ global annealing ansatz, where all nodes share the same scalar inverse temperature \(\beta^{(k)}\) at each layer \(k\), forming a schedule of total length \(p \in \{5, 10, 15\}\).

Sampling is performed on hardware using 720 Monte Carlo sweeps per layer. We choose a uniform initial distribution, achieved by appending \(\beta\) \,=\, $0$ to the beginning of the schedule. After sampling, configurations are transferred to the CPU where the CPU evaluates the average energy $\langle E \rangle$ using
Eq.~\eqref{eq3} and COBYLA updates the schedule $\beta^{(k)}$. In this
training loop the cost function is the average sampled energy,
$\langle E \rangle$, estimated from $N_E$ independent MCMC runs. After
training, however, each candidate schedule is evaluated on a fresh batch of
$10^5$ runs and assigned a success probability, defined as the fraction of runs
that reach the putative ground-state energy at the final layer. When referring
to the best schedule or best $\beta$ in
Fig.~\ref{fig:SA_discovery}, we mean the schedule that achieves the highest
success probability out of this independent batch. Insets in
Fig.~\ref{fig:SA_discovery}(c) show these success probabilities for all 100
optimizer runs, illustrating both the variability across runs and the
improvement with circuit depth.

The loop continues until either convergence or a maximum number of energy evaluations is reached. The algorithm is outlined in
Algorithm~\ref{alg:numerical beta-parameterized-PAOA}, and parameters are
listed in Table~\ref{tab2:simulation_parameters}.

\begin{figure*}[t!]
    \centering
    \includegraphics[width=1\linewidth]{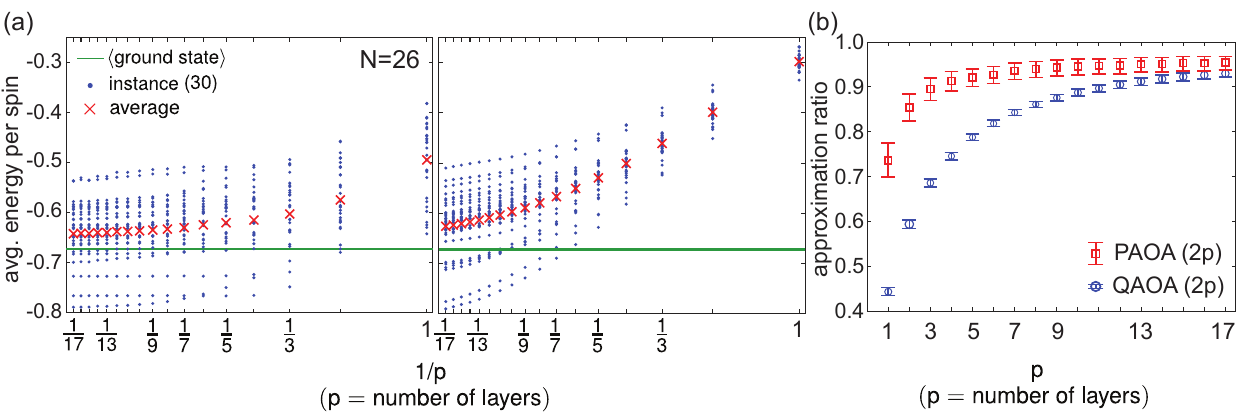}
    \caption{{\footnotesize \textbf{PAOA vs QAOA on the Sherrington-Kirkpatrick model}. (a)~PAOA results (left) using two-schedule ansatz ($\beta_1$ and $\beta_2$) with \(2p\) parameters compared against QAOA (right) with \(2p\) parameters ($\gamma$ and $\beta$). For each depth \(p\), the PAOA schedules are optimized on a separate training set; the average schedule is then applied to 30 random test instances of size $N$\,=\,$26$ without retraining. QAOA results use optimal parameters from prior work~\cite{farhi2022qaoa,QAOA_max_cut_farhi}. Red crosses denote averages across the 30 instances, blue dots show individual instance energies, and the solid green line indicates the average ground-state energy per spin. (b)~Approximation ratios of PAOA (red squares) and QAOA (blue circles), averaged across the 30 instances. Error bars indicate the 95\% confidence intervals computed from \(10^4\) bootstrap samples with replacement.}}
\label{fig:SK_Model}
\end{figure*}

The results of this optimization are summarized in Fig.~\ref{fig:SA_discovery}. The optimized schedules shift the energy distribution toward lower values (Fig.~\ref{fig:SA_discovery}b), increasing the probability of reaching the putative ground state at $E$\,=\,$-360$, which was independently found using a linear SA (from $\beta$\,=\,0.1 to 5) schedule with $10^6$ MCS. The schedules responsible for this improvement are shown in Fig.~\ref{fig:SA_discovery}c. Notably, without any constraints on monotonicity, PAOA consistently learns schedules that begin at high temperature (low \(\beta\)) and gradually cool (increase \(\beta\)), resembling classical SA despite starting from a flat $\beta$\,=\,$2$ schedule. 

As depth $p$ increases, our schedules both cool further and spend more total sweeps at lower temperatures, so the energy histograms naturally put more weight near the ground state. In majorization language, the deeper runs appear to majorize the shallower ones, but we only show energy marginals, therefore, this is suggestive rather than proof. Similar cooling-with-depth trends were observed for QAOA \cite{lotshaw2023approximate}. 

The elevated $\beta$ at the first step ($p\,$=1) for the 15-layer case may be a result of the substantial degeneracy among near-optimal schedules. Many distinct $\beta$ profiles achieve nearly identical success probabilities (shaded blue lines in Fig.~\ref{fig:SA_discovery}c). The inset histograms show that while individual schedules vary across 100 optimizer runs, performance remains stable (even for non-monotonic schedules) with the smallest variance in success probabilities for deeper schedules ($p=15$).

Constraints could be added to restrict the class of allowed schedules, e.g., enforcing linear, exponential, or monotonic cooling profiles using inequality constraints in COBYLA. We did not use such constraints  to avoid biasing the solution toward SA. These results demonstrate that SA emerges as a special case of variational Monte Carlo, recoverable from the global schedule ansatz. Moreover, PAOA proves to be more general, as it points to the possibility of alternative non-monotonic schedules with comparable performance, indicating the flexibility of the algorithm. 

\section{PAOA on FPGA with On-Chip Annealing}
\label{sec:FPGA_implementation}

To support long annealing schedules in 3D spin-glass problems, we designed a p-computer architecture that performs annealing entirely on-chip, unlike earlier implementations that require off-chip resources for annealing (e.g., \cite{aadit2022massively}). The annealing schedule $\beta^{({k})}$ is preloaded to the FPGA and indexed via a layer counter that advances after a fixed number of Monte Carlo sweeps (MCS). At each step, the current $\beta$ value is used to scale the synaptic input $I_i$ via a digital signal processing (DSP) multiplier (see Supplementary Fig.~\ref{fig:online_annealing}).

The full \(p\)-layer annealing process runs uninterrupted on the FPGA, and the final spin configurations are read out only after the last layer is complete.
\begin{table}[b!]
    \centering
    \caption{Wall-clock time for 3D spin glass ($L^3$\,=\,$6^3$), comparing CPU to FPGA with 10 identical replicas and $10^4$ independent runs.}
    \vspace{0.5pc}
    \label{tab:FPGA_advantage}
    \renewcommand{\arraystretch}{1}
    \setlength{\tabcolsep}{3pt}
    \begin{tabular}{@{}lccc|c@{}}
        \toprule
        Layers & 5 & 10 & 15 & Flips/ns \\
        Time (MCS/replica) & $3.6\times 10^7$ & $7.2\times 10^7$ & $1.08\times 10^8$ & \\
        CPU (s) & 1953 & 3800 & 5850 & 0.0398 \\
        FPGA-10 replicas (s) & 2.46 & 4.90 & 7.36 & 31.74 \\
        \bottomrule
    \end{tabular}
\end{table}

The FPGA operates in fixed-point arithmetic with \(s\{4\}\{5\}\) precision (where \(s\) denotes the sign bit, followed by 4 integer and 5 fractional bits) for both \(\beta^{(k)}\) and \(J_{ij}\). Due to the size of the FPGA we use, we instantiate 10 independent replicas of the $L^3$\,=\,$6^3$ spin-glass graph, enabling parallel updates of 2160 p-bits per cycle. Our FPGA architecture with on-chip annealing capability  achieves about  approximately an 800-fold reduction in wall-clock time in the FPGA compared to an optimized  CPU implementation with graph-colored Gibbs sampling (Table~\ref{tab:FPGA_advantage}).  

PAOA is parallelizable at multiple levels. First, the cost
estimator $\langle E\rangle$ is computed from $N_E$ independent MCMC runs, which
can be executed concurrently across replicas. Second, on sparse graphs, independent sets admit
parallelism via chromatic sampling. Consequently, high‑throughput
implementations are possible using CPUs, GPUs, FPGAs, custom accelerators or nanodevices. Our FPGA design is one point in this space and was chosen to accelerate the
experiments in this work. The wall‑clock numbers in Table~\ref{tab:FPGA_advantage}
are an implementation case study rather than a universal hardware comparison.

All simulations use a uniform initialization, implemented by prepending a $\beta$\,=\,0 layer to the schedule. This yields random initial conditions and ensures randomized spin states for each replica. Further hardware design details are provided in Supplementary Section 4.

\section{PAOA versus QAOA: Sherrington–Kirkpatrick Model}
\label{SK_model}
The Sherrington-Kirkpatrick (SK) model defines a mean-field spin glass with all-to-all random couplings, and serves as a canonical benchmark for optimization algorithms. The classical energy function is defined as
\begin{equation}
    E(\{m\}) = -\frac{1}{\sqrt{N}} \sum_{i<j} J_{ij} m_i m_j
    \label{sk-model}
\end{equation}
where \( m_i \in \{-1, 1\} \) and \( J_{ij} \sim \mathcal{N}(0,1) \). The exact ground-state energy per spin in the large-\(N\) limit is known~\cite{parisi1979infinite}. 

QAOA has been studied extensively on the SK model~\cite{farhi2022qaoa, QAOA_max_cut_farhi, crisanti2002analysis, gulbahar2023maximum}, with recent numerical work suggesting it can approximate the ground state in the infinite-size limit~\cite{boulebnane2025sk}. To enable a direct comparison, we evaluate PAOA using a two-schedule ansatz with exactly \(2p\) parameters, matching the parameter count of depth-\(p\) QAOA. Although PAOA scales readily to much larger $N$, we stick to a fixed size of $N$\,=\,$26$, a typically studied SK-model, to enable a direct comparison with QAOA.

Each schedule is assigned to one half of the graph, split uniformly at random. For $N$\,=\,$26$, we randomly generate 30 training instances and optimize PAOA separately for each instance and layer depth \(p \in \{1, \dots, 17\}\). The resulting schedules are then averaged and applied without further adjustment to a disjoint set of 30 test instances. QAOA results plotted in Fig.~\ref{fig:SK_Model}a use optimal parameters from prior work~\cite{farhi2022qaoa, QAOA_max_cut_farhi}. For each instance, exact ground states are computed by exhaustive search.

Our results show that PAOA has highly competitive performance under these iso-parametric conditions. As shown in Fig.~\ref{fig:SK_Model}a (left, PAOA), the average energy per spin consistently decreases with layer depth, and schedules trained on $N$\,=\,$26$ generalize well to much larger instances up to $N$\,=\,$500$ (see Supplementary Fig.~\ref{fig:Large_SK_PAOA}). To formalize this comparison, we report the approximation ratio in Fig.~\ref{fig:SK_Model}b, defined as
\begin{equation}
    \text{Approx. Ratio} = \frac{E(\{m\})}{E(\{m_{\mathrm{sol}}\})}\,
    \label{eq:approx_ratio}
\end{equation}
where \(E(\{m_{\mathrm{sol}}\})\) is the cost of the exact ground state. PAOA achieves a consistently higher approximation ratio than standard QAOA across all depths. Notably, the performance of vanilla PAOA is comparable to that of recently proposed hybrid quantum-classical methods that use classical techniques to improve QAOA's performance~\cite{QRR_QAOA}, suggesting that a definitive quantum advantage on this problem has yet to be firmly established.

We emphasize that a clear quantum advantage over classical algorithms is expected to emerge in problems where interference plays an explicit role, such as in the synthetic examples constructed by Montanaro et al.~\cite{montanaro2024speedup}. In these cases, the superposition of many computational paths, enabled by either the problem structure or the quantum algorithm, creates interference patterns that amplify the correct solution while suppressing others, an effect that classical probabilistic samplers struggle to reproduce due to the well‑known ``sign problem''~\cite{troyer2005computational}. As shown in~\cite{chowdhury2023emulating}, a probabilistic formulation of generic unitary evolutions yields complex (often purely imaginary) weights, so sampling  effectively becomes randomized. While post‑phase corrections recover unbiased estimates, the number of samples grows exponentially.

These interference‑dominated regimes are precisely where we do not expect PAOA, or any classical sampler, to compete. On the other hand, unless such interference‑driven advantages are clearly identified and demonstrated more broadly across practical problem classes, we do not expect any significant quantum advantage from QAOA.  Meanwhile, PAOA offers a rigorous, scalable and high-performing benchmark for variational sampling at problem sizes  beyond the reach of current quantum devices.

\begin{figure*}[t!]
    \centering
    \includegraphics[width=\linewidth]{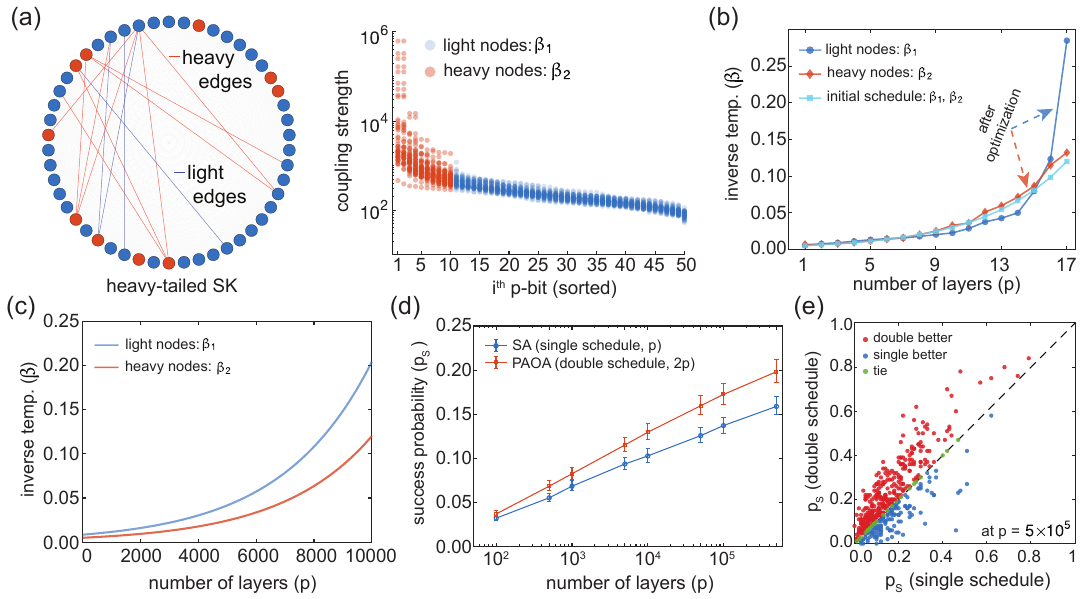}
    \caption{{\footnotesize
    \textbf{Learning a variational principle using a PAOA double‑schedule ansatz}.
(a) Heavy‑tailed SK: per‑node coupling strengths for $50$ instances of \(N{=}50\), sorted in descending order; the heavy‑tailed distribution separates heavy (assigned $\beta_2$ schedule) and light (assigned $\beta_1$ schedule) nodes.
(b) PAOA training with two schedules (\(2p\) parameters) assigned to heavy and light nodes based on their coupling strengths, initialized from an optimized single annealing schedule (cyan). Curves are averaged across 50 instances; red/blue denote heavy/light nodes.
(c) The extrapolated double schedule suggested by (b): the heavy‑node schedule is extended to more layers, and the light‑node schedule is scaled up following PAOA’s guidance.
(d) Average success probability over 500 instances comparing single‑schedule SA (blue) and double‑schedule PAOA (red) for \(N\,{=}\,50\). Error bars are 95\% confidence intervals from \(10^5\) bootstrap samples with replacement. (e) Success probability showing per-instance comparison between PAOA ($2p$) and SA ($p$) at \(p=5\times 10^5\).
    }}
    \label{fig:SK_Levy_Model}
\end{figure*}
\section{Multi-Schedule Annealing in the L\'evy SK Model}
\label{sec:SK_model_with_Levy_bonds}
To a heuristic solver, the SK model presents no intrinsic structure beyond a dense random coupling graph. To explore whether PAOA can adaptively exploit structure when it might exist, we consider a variant where the coupling weights $J_{ij}$ follow a heavy-tailed L\'evy distribution~\cite{boettcher2012ground}:
\begin{equation}
    P(J) = \frac{\alpha}{2} |J|^{-1-\alpha}, \qquad |J| > 1
\end{equation}
We use $\alpha$\,=\,$0.9$, which produces a distribution with diverging variance, dominated by rare, large couplings.

In this regime, a natural question arises: can PAOA discover principles that treat strongly and weakly coupled nodes differently? Prior work on Boltzmann machines and annealing schedules~\cite{aarts2005simulated, adame2020inhomogeneous} suggests that high-degree or strongly coupled nodes benefit from slower annealing (more time for high temperature) to avoid freezing too early. In these prior works, the heuristics are hand-designed by the ingenuity of the human algorithm designers.  Here, we test whether the principle of annealing subgraphs with different temperatures can be discovered by PAOA through black-box optimization. 

We begin by introducing a coupling strength per node: 
\begin{equation}
    \lambda_i = \sum_j |J_{ij}|
    \label{coupling_strength}
\end{equation}

We then sort nodes by  \(\lambda_i\) and split into a heavy group (top 20\%) and a light group (bottom 80\%), as shown in Fig.~\ref{fig:SK_Levy_Model}a. The split here is arbitrary and we confirmed that a 50\%-50\% partition following the same procedure yields similar results (see Supplementary Fig.~\ref{fig:two_schedules_50_50}). 

To test if PAOA can leverage this structure, we first identify a near-optimal single-schedule using simulated annealing at shallow depth ($p$\,=\,$17$). We then let COBYLA optimize two-schedules, corresponding to heavy and light nodes, freely. After convergence, we average the schedules across all instances. As shown in Fig.~\ref{fig:SK_Levy_Model}b, the optimized schedules exhibit a qualitative separation: heavy nodes are assigned a higher-temperature (lower $\beta$) profile, and light nodes anneal faster. We verified that the different schedules for heavy and light nodes  lead to higher success probability for the double-schedule over the single-schedule. Due to the shallow depth ($p$\,=\,$17$) however the differences are not appreciably large. 

However, when extrapolating the double-schedules discovered by PAOA to much deeper layers (Fig.~\ref{fig:SK_Levy_Model}c), we observe a clear difference against single-schedule SA. Our extrapolation is relatively straightforward but we report the details in the Methods Section. The results, shown in Fig.~\ref{fig:SK_Levy_Model}d for $N$\,=\,$50$ and 500 unseen instances, show that the two-schedule ansatz outperforms the single schedule, with a clear gap in success probabilities between the two approaches, confirming a robust benefit from structured, heterogeneous annealing. Details on determining the success probability for single and double schedules are in the Methods Section.   

This behavior suggests that PAOA can discover problem-aware heuristics computationally. While such adaptive schedules have long been used manually in structured settings \cite{mohseni2021nonequilibrium}, we are not aware of prior variational algorithms for non-equilibrium sampling that systematically learn them from data. In this sense, PAOA serves as a tool for discovering new algorithmic strategies in disordered systems.

\section{Outlook}
\label{sec:conclusions}
We have introduced the Probabilistic Approximate Optimization Algorithm (PAOA), a variational Monte Carlo framework that generalizes simulated annealing and supports a wide range of parameterizations compatible with existing Ising hardware. By treating the energy landscape itself as a variational object and updating it through classical feedback from Monte Carlo samples, PAOA enables efficient sampling from non-equilibrium distributions using shallow Markov chains.

We demonstrated that PAOA captures simulated annealing as a limiting case and, without constraints, can rediscover SA-like behavior when optimized for energy minimization. Using a minimal global schedule ansatz, PAOA efficiently learned monotonic cooling profiles that converge to low-energy states on 3D spin-glass problems. The approach was implemented on an FPGA-based p-computer, achieving hardware-accelerated annealing at rates exceeding CPU-based implementations by several orders of magnitude.

Beyond global schedules, we explored richer ansätze with multiple temperature profiles. On the Sherrington-Kirkpatrick (SK) model, PAOA exceeded the performance of QAOA under iso-parametric conditions, while scaling to system sizes beyond those accessible to quantum devices. In a heavy-tailed variant of the SK model, PAOA automatically learned to assign slower annealing schedules to strongly coupled nodes, reproducing hand-crafted heuristics from prior work. As such, PAOA can be a  tool for discovering new algorithmic strategies for sampling in disordered systems.

\section{Methods}
Numerical simulation uses double precision (64-bit) in C++ to generate results for PAOA in Fig.~\ref{fig:majority_gate}d, Fig.~\ref{fig:SK_Model}a, and Supplementary Fig.~\ref{fig:full_adder_sequential_update}d. QAOA results were reproduced from~\cite{lykov2023fast}.

\subsection*{Data transfer between FPGA and CPU}
\label{sec:data_transfer}
A PCIe interface was used to communicate between FPGA and CPU
through a MATLAB interface for the ‘read/write’ operations. A global ‘disable/enable’ signal broadcast from MATLAB to the FPGA was used to freeze and resume all p-bits. Before a ‘read’ instruction, the p-bit states were saved to the local block memory (BRAM) with a snapshot signal. Then the data were read once from the BRAM using the PCIe interface and sent to MATLAB for post-processing, that is, updating the annealing schedule ($\beta^{(k)}$). For the ‘write’ instruction, the ‘disable’ signal was sent from MATLAB to freeze the p-bits before sending the updated schedule. After the ‘write’ instruction was given, p-bits were enabled again with the ‘enable’ signal sent from MATLAB.  

\subsection*{Measurement of wall-clock time and flips per nanosecond}
To measure the annealing wall-clock time per replica in FPGA along with the reading and writing overhead, we use MATLAB’s built-in tic and toc functions and average the total annealing time over 100 independent experiments. Adding all flip attempts for the entire system with all replicas, 2160 p-bits, the total flips per nanosecond (flips/ns) are computed. For CPU measurements, MATLAB’s built-in tic and toc functions were used to measure the total annealing time taken to perform the MCS in Table~\ref{tab:FPGA_advantage} with 1000 runs each. The average time per MCS is reported for a single replica for a network size of $L^3$\,=\,$6^3$. 

\subsection*{Schedule Generation for the L\'evy SK Model Benchmark}

The performance benchmark in Section \ref{sec:SK_model_with_Levy_bonds} required comparing two annealing strategies. The schedules for these strategies were generated via a grid search based on a geometric functional form:
\begin{equation}
    \beta^{(k)} = \beta_\text{initial} \cdot \left(\frac{\beta_{\text{final}}}{\beta_{\text{initial}}}\right)^{k/(p-1)}, \quad k = 0, \dots, p-1
    \label{geometric_schedule_methods}
\end{equation}
The parameter ranges searched were \(\beta \in [0.005,\ 0.12]\) for the $N$\,=\,$50$ instances.

The two strategies were constructed as follows:
\begin{itemize}
    \item Single-Schedule (SA): We performed a grid search over \(\beta_\text{final}\) to find the single geometric schedule that produced the lowest average energy across all instances. This served as the baseline.
    \item Two-Schedule PAOA: The optimized baseline schedule was assigned to the heavy nodes, as suggested by PAOA. For light nodes, we scale up the baseline schedule by a scaling factor that reflects the separation proposed by PAOA. Then, both geometric schedules are extrapolated to large number of layers to study the effect of deep PAOA. 
\end{itemize}

In order to compare the two approaches, we define success probability as the ratio of the number of runs that hit the putative ground state to the total number of runs (i.e., 100 independent runs). The putative ground state was found by running a long simulated annealing with one million sweeps and ten independent runs; then choosing the lowest energy among all ten million states.       

\subsection*{Optimizer choice}
For the outer optimization loop, we used the COBYLA algorithm as it is a robust and widely used method for derivative-free optimization problems with a moderate number of variables \cite{pellow2021vqls, arrasmith2021effect, schiffer2022adiabatic}. Our COBYLA implementation is based on the the non-linear optimization package (Nlopt) \cite{NLopt}. While the performance of PAOA may depend on the choice of classical optimizer, a detailed comparison of different optimization techniques is beyond the scope of this work.

\section*{Data Availability}
All generated and processed data used in this study are openly accessible and can be found in the Github repository \cite{Abdelrahman2025PAOA}. Other findings of this study are available from the corresponding authors upon request.

\section*{Code Availability}
The MATLAB and C++ implementations of PAOA used to generate the results and plots in this study can be found in the Github repository mentioned in the data availability section.

\subsection*{Acknowledgements}
ASA, SC, and KYC acknowledge support from the National Science Foundation (NSF) under award number 2311295, and the Office of Naval Research (ONR), Multidisciplinary University Research Initiative (MURI) under Grant No. N000142312708. FM was partially supported by the Office of Naval Research (ONR) under Award No. N00014-23-1-2771. We are grateful to Navid Anjum Aadit for  discussions related to the hardware implementation of online annealing, and Ruslan Shaydulin and  Zichang He for input on QAOA benchmarking. Use was made of computational facilities purchased with funds from the National Science Foundation (CNS-1725797) and administered by the Center for Scientific Computing (CSC). The CSC is supported by the California NanoSystems Institute and the Materials Research Science and Engineering Center (MRSEC; NSF DMR 2308708) at UC Santa Barbara.

\section{Author contributions}
ASA and KYC conceived the study. ASA led the full implementation of PAOA, including all simulations and experiments, building on the initial PAOA implementation by SC that used black-box optimizers. ASA designed and built the FPGA implementation of PAOA with on-chip annealing capability.  SC performed the QAOA benchmarking. FM contributed key theoretical insights and assisted in interpreting the spin-glass benchmarks. All authors contributed to discussion of the results and writing of the manuscript.

\section{Competing interests}
The authors declare no competing interests.

\clearpage

\onecolumngrid

\begin{center}
{\sffamily\Large\bf Supplementary Information\par}
\vskip 0.5em
{\sffamily\LARGE\bf Probabilistic Approximate Optimization: A New Variational Monte Carlo Algorithm \par}
\vspace{1em}
\normalfont\noindent{\sffamily Abdelrahman S.  Abdelrahman, Shuvro Chowdhury, Flaviano Morone  and Kerem Y. Camsari}
\end{center}

\beginsupplement
In the main manuscript, we discussed privatized-beta-based PAOA, and global-beta-based PAOA. Here, we present the implementation of fully parameterized PAOA (See Algorithm.~\ref{alg:numerical_PAOA_learning}), which is used in learning the full-adder target states by minimizing the negative log-likelihood cost (expressed in equation \eqref{algo1_eq4}), see Section.~\ref{sequential_update_FA}. The exact same structure of the algorithm is also used for learning the majority gate correct states, however, with private $\beta$ as ansatz, see Section.~\ref{MAJ_gate_extended}.  
\begin{figure}[htp!]
    \centering
    \begin{algorithm}[H]
    \DontPrintSemicolon
\SetKwInOut{Input}{Input}
\SetKwInOut{Output}{Output}
\caption{PAOA: fully-parameterized}
\label{alg:numerical_PAOA_learning}
\Input{number of nodes $N$, number of layers $p$, number of experiments $N_E$, initial variational parameters $(J^{(1)}, J^{(2)},\dots,  J^{(p)})$, tolerance $\varepsilon_{\text{step}}$, maximum iterations $t_{\max}$, truth table}
\Output{trained set of weights $(J^{(1)}_{\text{opt}}, J^{(2)}_{\text{opt}},\dots,  J^{(p)}_{\text{opt}})$}
\SetKwFunction{FsubMain}{p-computer}
\SetKwFunction{FMain}{PAOA-circuit}
\SetKwProg{Fn}{Function}{:}{}

\Fn{\FsubMain{$J_{\text{init}}$, N, p}}{
    initialize all spins randomly\;
    \For{$i \gets 1$ \KwTo $p$}{
        $J \gets J^{(i)}$\;
        \For{$j \gets 1$ \KwTo $N$}{
            solve equations \eqref{algo1_eq1} and \eqref{algo1_eq2}
        }
    }
    
    \Return{p-bit states in decimal}
    }
    \Fn{\FMain{$N_E$, $J$, N, p}}{
        \For{$k \gets 1$ \KwTo $N_E$}{
        state $\gets$ \FsubMain{$J$, N, p}\;
        save the p-bit states\;
        }
        find the estimated distribution ($\hat{\boldsymbol{\rho}}_p$) using equation \eqref{algo1_eq3}\;
        compute the cost ($\mathcal{L}$) using equation \eqref{algo1_eq4}\;
        \Return{cost}
        }
        
        \While{({ step size $>\varepsilon_{\text{step}}$ and number of iterations  $< t_{\max}$})}{
        cost $\gets$ PAOA-circuit($N_E$, $J$, N, p)\;
        minimize cost and get a perturbation vector ($p$) using a gradient-free optimizer\;
        \For{$i \gets 1$ \KwTo $p$}{
        $J^{(i)}_{t+1} \gets J^{(i)}_{t} + p^{(i)}$\;
    }
   $t\gets t+1$\;
        }    
    
    \Return{optimal variational parameters}

\end{algorithm}
\end{figure}
The p-computer subroutine in Algorithm.S1 uses p-bit equations to generate samples from a ($J,h$) parameterized distribution. In equations.~\eqref{algo1_eq1} and \eqref{algo1_eq2}, we used two variations of this algorithm. In the full-adder case, we set $\beta$\,=\,$1$, $h$\,=\,$0$, and optimize $J$. 
\begin{equation}
    m_i = \text{sgn}[\tanh(\beta I_i) - \text{rand}_{\text{u}}(-1, 1)],
    \label{algo1_eq1}
\end{equation}
In the majority gate problem, the graph weights and biases are set to one and zero, respectively ($J_{ij}=+1, h=0$). Then, the annealing schedule is localized for each node ($\beta_i$), that is, each p-bit gets its own schedule. 
\begin{equation}
    I_i = \sum_j J_{ij} m_j + h_i.
    \label{algo1_eq2}
\end{equation}
Since the cost function is defined over the distribution of the target states, we estimated the distribution using the generated independent samples by counting the frequency of observing the desired states over $N_E$ independent experiments (Eq.~\eqref{algo1_eq3}). For accurate estimation, we use ten million experiments, which bounds the deviation from the true distribution to approximately $3.2\times 10^{-4}$. 
\begin{equation}
    \hat{\boldsymbol{\rho}}^{(j)}(\{m\}) = \frac{1}{N_E} \sum_{k=1}^{N_E} \mathbbm{1}\{X_{j,k} = \{m\}\}, \quad \{m\} \in \Omega
    \label{algo1_eq3}
\end{equation}
In the negative log-likelihood function defined in Eq.~\eqref{algo1_eq4}, the estimated distribution over a target states set $\mathcal{X}$ is used to calculate the associated loss. The variational parameters $\boldsymbol{\theta}$ represent the parameters being optimized, such as $J_{ij}$ for the full-adder, and $\beta_i$ for the majority gate problem. 
\begin{equation}
    \mathcal{L}(\boldsymbol{\theta}) = -\sum_{\{m\} \in \mathcal{X}} \ln(\hat{\boldsymbol{\rho}}_p(\{m\}; \boldsymbol{\theta})),
    \label{algo1_eq4}
\end{equation}

\setcounter{secnumdepth}{1}
\renewcommand{\thesection}{\arabic{section}}
\section{Full-Adder with Fully Parametrized PAOA Ansatz}

\label{sequential_update_FA}
\begin{figure}[htp!]
    \centering
    \includegraphics[width=1\linewidth]{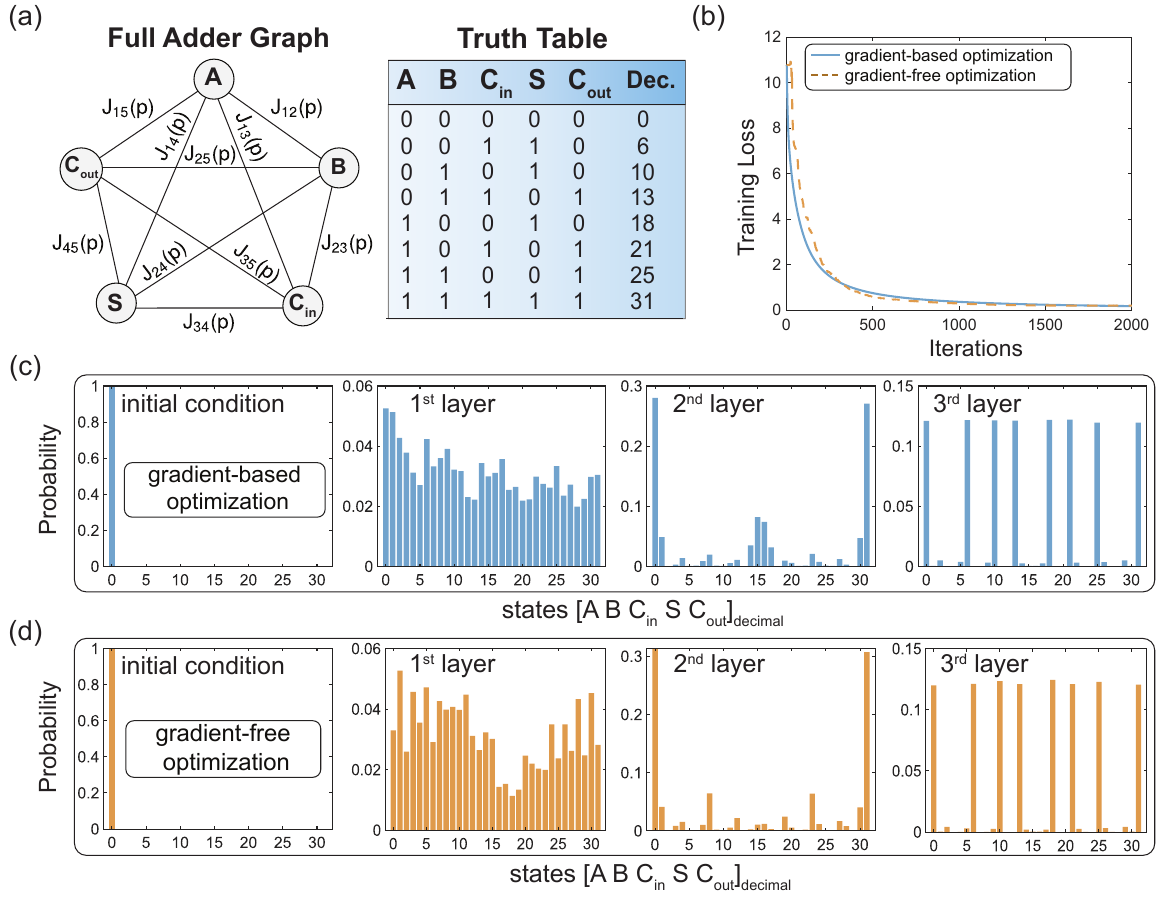}
    \caption{(a) All-to-All full-adder network, showing  graph weights $ J(p)$, and truth table, where Dec. refers to the decimal representation of the state of [A B C$_{\text{in}}$ S C$_{\text{out}}$] from left to right. (b) Training loss over optimization iterations for gradient-based and gradient-free methods.  (c) PDF evolution across three layers, obtained from 
    a Markov chain. (d) PDF evolution across three time layers, using MCMC ($10^{7}$ experiments). }
    \label{fig:full_adder_sequential_update}
\end{figure}
The full-adder is represented as a fully connected network comprising five p-bits, where the graph weights are the only variational parameters (as shown in Supplementary {Fig.~\ref{fig:full_adder_sequential_update}a}). As depicted in the same figure, the graph weights evolve with $p$, such that for each layer, we have different weights that are responsible for the PDF evolution. The full-adder performs 1-bit binary addition with three inputs (A, B, and Carry in = \( \text{C}_{\text{in}}\)) and two outputs (Sum = S and Carry out = \( \text{C}_{\text{out}}\)). We use a finite Markov chain with three layers ($p$=$3$), and $\boldsymbol{\rho}_0  = \begin{bmatrix}
    1 & 0 & 0 & 0 & 0
\end{bmatrix}^\top$ as the initial condition PDF. The optimization is carried out exactly similar to the majority gate problem in Section.~\ref{MAJ_gate_extended}, except the parameterization here is the graph weights. The optimal parameters, rounded to the nearest hundredth, are presented in Eq.~\eqref{FA_params}. The results in Supplementary Fig.~\ref{fig:full_adder_sequential_update}c and Fig.~\ref{fig:full_adder_sequential_update}d show clear agreement between the two methods, namely, the gradient-based approach where the probability transition matrix is constructed, and the gradient-free one where samples are generated to estimate the underlying distribution. More details on the gradient-based method are provided in Section.~\ref{ANDGate}, where we show how an AND gate can be solved analytically.  

{\scriptsize{\begin{equation}
    \begin{aligned}
        &J^{(1)}_{\text{optimal}} \text{=} \begin{pmatrix}
0 & 0.18 & 0.12 & 0 & -0.10 \\
0.18 & 0 & -0.08 & 0.06 & -0.11 \\
0.12 & -0.08 & 0 & 0.01 & -0.05 \\
0 & 0.06 & 0.01 & 0 & 0.02 \\
-0.10 & -0.11 & -0.05 & 0.02 & 0
\end{pmatrix},  J^{(2)}_{\text{optimal}}  \text{=} \begin{pmatrix}
0 & 0.31 & 0.43 & 1.14 & 0.17 \\
0.31 & 0 & 0.15 & 0.58 & -0.32 \\
0.43 & 0.15 & 0 & 1.26 & -0.28 \\
1.14 & 0.58 & 1.26 & 0 & 1.45 \\
0.17 & -0.32 & -0.28 & 1.45 & 0
\end{pmatrix}, J^{(3)}_{\text{optimal}}  \text{=} \begin{pmatrix}
0 & -1.68 & -1.91 & 1.90 & 1.93 \\
-1.68 & 0 & -2.31 & 1.87 & 2.25 \\
-1.91 & -2.31 & 0 & 1.84 & 2.58 \\
1.90 & 1.87 & 1.84 & 0 & -3.93 \\
1.93 & 2.25 & 2.58 & -3.93 & 0
\end{pmatrix}
    \end{aligned}
    \label{FA_params}
\end{equation}
}	}
The simulation parameters used here are tabulated in Table~\ref{tab:simulation_parameters}. In gradient-based approach, the training terminates when either the tolerance in gradient, $\|\nabla \mathcal{L}(\boldsymbol{\theta})\|<\varepsilon_{\text{grad}},$ or maximum iterations is met because updates of such order no longer improve the objective. For derivative‑free (Cobyla), convergence is declared when the
step‑size (trust–region radius) falls below a chosen tolerance $\Delta \theta < \varepsilon_{\text{step}},$ signalling that all admissible simplex moves would alter the variational parameters by a negligible change, and hence further iterations are unproductive.

\begin{table}[htp!]
    \centering
    \caption{Full Adder Simulation Parameters.}
        \vspace{0.5pc}
    \label{tab:simulation_parameters}
    \renewcommand{\arraystretch}{1.2} 
    \setlength{\tabcolsep}{8pt} 
    \begin{tabular}{@{}ll@{}}
        \toprule
        Parameter & Value \\ 
        \midrule
        number of nodes ($N$) & 5  \\ 
        number of layers ($p$) & 3  \\ 
        update order & $\{m_1, m_2, m_3, m_4, m_5\}$\\ 
        initial parameters ($J^{(1)}, J^{(2)}, J^{(3)}$) & $0.1$  \\ 
        learning rate ($\eta$) & $0.01$ \\ 
        tolerance in gradient ($\varepsilon_{\text{grad}}$) & $10^{-6}$  \\ 
        maximum iterations ($t_{\max}$) & $2000$ \\ 
        \midrule
        number of experiments ($N_E$) & $10^7$ \\ 
            tolerance ($\varepsilon_{\text{step}}$) & $10^{-6}$ \\ 
        \bottomrule
\end{tabular}
\end{table}  
\section{Majority Gate Problem}
\label{MAJ_gate_extended}
In this section, we present the optimal parameters obtained for the majority gate problem, solved in the Representative problem: majority gate subsection in the main text, using a local-annealing schedule ansätze with two layers ($p$\,=\,$2$). In this formulation, each node is assigned a private inverse temperature $\beta$, allowing the model to adaptively capture local structure in the optimization landscape. The full set of simulation parameters used to obtain these results is summarized in Table~\ref{tab:simulation_parameters_MAJ}.
\begin{table}[htp!]
    \centering
    \caption{Majority Gate Simulation Parameters.}
        \vspace{0.5pc}
    \label{tab:simulation_parameters_MAJ}
    \renewcommand{\arraystretch}{1.2} 
    \setlength{\tabcolsep}{8pt} 
    \begin{tabular}{@{}ll@{}}
        \toprule
        Parameter & Value \\ 
        \midrule
        number of nodes ($N$) & 4  \\ 
        number of layers ($p$) & 2  \\ 
        update order & $\{m_1, m_2, m_3, m_4\}$\\ 
        initial parameters ($\bar{J}^{(1)}, \bar{J}^{(2)}$) & $1$  \\ 
        learning rate ($\eta$) & $0.004$ \\ 
        tolerance in gradient ($\varepsilon_{\text{grad}}$) & $10^{-7}$  \\ 
        maximum iterations ($t_{\max}$) & $5000$ \\ 
        \midrule
        number of experiments ($N_E$) & $10^7$ \\ 
         tolerance ($\varepsilon_{\text{step}}$) & $10^{-7}$ \\ 
        \bottomrule
    \end{tabular}
    \end{table}    

The optimal parameters for the nodes labeled \( [A, B, C, Y] \), rounded to the nearest decimal, are:
\begin{equation}
    \begin{blockarray}{ccc}
        & p=1 & p=2 \\  
      \begin{block}{c[cc]}  
         \beta_A  & 0.8 & -0.2   \\
          \beta_B &  0.8  & 0    \\
          \beta_C &  0.9    & 0    \\
          \beta_Y &   0.5  & 2.7 \\
        \end{block}
      \end{blockarray}
\end{equation}
These variational parameters, when combined with the initial graph weights ($J_{ij}$\,=\,$+1$), define the effective couplings $\bar{J}^{(k)}_{ij}$\,=\,$\beta^{(k)}_iJ_{ij}$. The optimized parameters can be used to solve the majority gate problem (shown in Figs.~\ref{fig:majority_gate} in the main text).

\section{Analytical Formulation of PAOA: AND Gate Problem}
\label{ANDGate}
\begin{figure}[htp!]
\vspace{-1pc}
    \centering
    
    \begin{tikzpicture}
    \node at (-0.5, 2.2) {\large (a)};

    \draw[thick] (0,0) circle (0.5cm);
    \node at (0,0) {\large $m_2$};

    \draw[thick] (3,0) circle (0.5cm);
    \node at (3,0) {\large $m_3$};

    \draw[thick] (1.5,2) circle (0.5cm);
    \node at (1.5,2) {\large $m_1$};

     \draw[thick, ->] (0.5,2) -- (1,2);
    \node at (0.3,2) {\large $h_1$};

    \draw[thick, ->] (-1,0) -- (-0.5,0);
    \node at (-1.25,0) {\large $h_2$};

    \draw[thick, ->] (4,0) -- (3.5,0);
    \node at (4.3,0) {\large $h_3$};

    \draw[thick, -] (0.5,0) -- (2.5,0);

    \draw[thick, -] (1.4/4,1.4/4) -- (1.45/1.23,2/1.23);

    \draw[thick, -] (1.5+1.4/4,2-1.4/4) -- (3-1.4/4,1.4/4);

    \node at (0.4,1.2) {\large $J_{12}$};
    \node at (1.5,0.3) {\large $J_{23}$};
    \node at (2.6,1.2) {\large $J_{13}$};

    \node (and) [and gate US, draw, logic gate inputs=nn, scale=3] at (7,1)  {};
    
    \draw ([xshift=-0.5cm] and.input 1) -- (and.input 1) node[midway, left=7pt] {\large \(m_1\)};
    \draw ([xshift=-0.5cm] and.input 2) -- (and.input 2) node[midway,  left=7pt] {\large \(m_2\)};
    \node at (5, 2.2) {\large (b)};

    \draw (and.output) -- ([xshift=0.5cm] and.output) node[midway,  right=7pt] {\large \(m_3\)};

    \node at (7.2,-0.2) {\large $m_1 \cap m_2  =m_3$};

    \node at (11,2.7) {\large \text{Truth Table}};
    \node at (11,1) {  
        \renewcommand{\arraystretch}{1.3}  
        \begin{tabular}{cccc} 
            \toprule
            \large $m_1$ & \large $m_2$ & \large $m_3$ & \large \text{Dec.} \\ 
            \midrule
            \large 0 & \large 0 & \large 0 & \large 0  \\ 
            \large 0 & \large 1 & \large 0 & \large 2  \\ 
            \large 1 & \large 0 & \large 0 & \large 4  \\ 
            \large 1 & \large 1 & \large 1 & \large 7  \\ 
            \bottomrule
        \end{tabular}
    };
    
\end{tikzpicture}

\vspace{-0.5pc}
    \caption{(a) A schematic of AND gate as an undirected graph. (b) The AND gate schematic along with the truth table, where Dec. refers to the decimal representation of the state [$m_1$,$m_2$,$m_3$] from left to right.}
    \label{fig:AND_gate}
\end{figure}

In this section, we show the details of learning the weights of an AND gate analytically using the fully parameterized PAOA ansatz. The problem is shown in Supplementary Fig.~\ref{fig:AND_gate}. Using the fully-parametrized-PAOA, the $J$ graph weight matrix and bias vector $h$ can be written as:
\begin{equation}
    J = \begin{pmatrix}
    0 & J_{12} & J_{13} \\
    J_{12} & 0 & J_{23} \\
    J_{13} & J_{23} & 0 \\    
    \end{pmatrix}, \quad h = \begin{bmatrix}
    h_1 \\
    h_2 \\
    h_3 \\    
    \end{bmatrix}
    \label{jand}
\end{equation}

For a general graph, the entries of \(w_k\) are given by
\begin{equation}
    \begin{aligned}
         &[w_k]_{ab} = P(a\leftarrow b)\\  
                    &= \begin{cases}
            \displaystyle \frac{1+m^{(a)}_k\tanh (I^{(b)}_k)}{2} &  \text{\footnotesize{if $m^{(a)}$ and $m^{(b)}$ are identical}}\\
             &   \text{\footnotesize{except possibly at the $k^{\rm th}$ bit}},\\
            0 &  \text{\footnotesize{otherwise}}
        \end{cases}
    \end{aligned}
        \label{single_update_W_matrix}
    \end{equation}
    
where \(I_k^{(b)} = \sum_{\ell} J_{k\ell} m_\ell^{(b)} + h_k\) is the synaptic input to node \(k\) based on the configuration \(m^{(b)}\), and \(m^{(a)}\) denotes the spin configuration after updating bit \(k\). 

Using equations \eqref{jand} and \eqref{single_update_W_matrix} with the following update order $\{m_1, m_2, m_3\}$,  the probability transition matrix ($W$) can be constructed as follows:

\begin{equation}
\begin{aligned}
    W &= w_3w_2w_1\\
    &=\underbrace{\begin{blockarray}{cccccccc}
\begin{block}{[cccccccc]}       
  t & t & 0 & 0 & 0 & 0 & 0 & 0 \\
 t' & t' & 0 & 0 & 0 & 0 & 0 & 0 \\
 0 & 0 & u & u & 0 & 0 & 0 & 0 \\
 0 & 0 & u' & u' & 0 & 0 & 0 & 0 \\
 0 & 0 & 0 & 0 & u' & u' & 0 & 0 \\
 0 & 0 & 0 & 0 & u & u & 0 & 0 \\
 0 & 0 & 0 & 0 & 0 & 0 & t' & t' \\
 0 & 0 & 0 & 0 & 0 & 0 & t & t \\
  \end{block}
\end{blockarray}}_{\text{update } m_3 \text{ conditioned on } \{m_2, m_1\}}\times \underbrace{\begin{blockarray}{cccccccc}
\begin{block}{[cccccccc]}      
  r & 0 & r & 0 & 0 & 0 & 0 & 0 \\
 0 & s & 0 & s & 0 & 0 & 0 & 0 \\
 r' & 0 & r' & 0 & 0 & 0 & 0 & 0 \\
 0 & s' & 0 & s' & 0 & 0 & 0 & 0 \\
 0 & 0 & 0 & 0 & s' & 0 & s' & 0 \\
 0 & 0 & 0 & 0 & 0 & r' & 0 & r' \\
 0 & 0 & 0 & 0 & s & 0 & s & 0 \\
 0 & 0 & 0 & 0 & 0 & r & 0 & r \\
  \end{block}
\end{blockarray}}_{\text{update } m_2 \text{ conditioned on } \{m_3, m_1\}}\times \underbrace{\begin{blockarray}{cccccccc}
\begin{block}{[cccccccc]}      
   p & 0      & 0        & 0  & p        & 0     & 0    & 0 \\
     0    & q     & 0      & 0    & 0      & q    & 0   & 0 \\
     0    & 0     & q'      & 0     & 0     & 0    & q'   & 0 \\
            0  & 0    & 0     & p'     & 0   & 0   & 0  & p'\\
           p'  & 0   & 0    & 0     & p'    & 0   & 0   & 0\\
           0   & q'    & 0     & 0     & 0     & q'     & 0    & 0\\
           0   & 0    & q     & 0     & 0    &       & q    & 0\\
           0    & 0     & 0        & p     & 0      & 0     & 0     & p\\
  \end{block}
\end{blockarray}}_{\text{update } m_1 \text{ conditioned on } \{m_3, m_2\}}\\
    &=\begin{blockarray}{ccccccccc}
  & 000 & 001 & 010 & 011 & 100 & 101 & 110 & 111 \\ 
\begin{block}{c[cccccccc]}      
   000  & prt & qst      & q'rt        & p'st   & prt        & qst     & q'rt    & p'st \\
    001 &  prt'    & qst'     & q'rt'      & p'st'    & prt'      & qst'    & q'rt'   & p'st' \\
    010 &  pr'u    & qs'u     & q'r'u      & p's'u     & pr'u     & qs'u    & q'r'u   & p's'u \\
    011 &         pr'u'  & qs'u'    & q'r'u'     & p's'u'     & pr'u'   & qs'u'   & q'r'u'  & p's'u'\\
    100 &        p's'u'  & q'r'u'   & qs'u'    & pr'u'     & p's'u'    & q'r'u'   & qs'u'   & pr'u'\\
    101 &        p's'u   & q'r'u    & qs'u     & pr'u     & p's'u     & q'r'u     & qs'u    & pr'u\\
    110 &        p'st'   & q'rt'    & qst'     & prt'     & p'st'    &  q'rt'     & qst'    & prt'\\
    111 &        p'st    & q'rt     & qst        & prt     & p'st      & q'rt     & qst     & prt\\
  \end{block}
\end{blockarray}
\end{aligned}
\end{equation}

where
\begin{equation}
    \begin{aligned}
        &p = \frac{1+\tanh(J_{12}+J_{13}+h_1)}{2}, \quad q= \frac{1+\tanh(J_{12}-J_{13}+h_1)}{2}, \quad  r= \frac{1+\tanh(J_{12}+J_{23}+h_2)}{2}\\
        &p' = \frac{1-\tanh(J_{12}+J_{13}+h_1)}{2}, \quad q'= \frac{1-\tanh(J_{12}-J_{13}+h_1)}{2}, \quad r'= \frac{1-\tanh(J_{12}+J_{23}+h_2)}{2}\\
        &s= \frac{1+\tanh(J_{12}-J_{23}+h_2)}{2}, \quad t = \frac{1+\tanh(J_{13}+J_{23}+h_3)}{2}, \quad u = \frac{1+\tanh(J_{13}-J_{23}+h_3)}{2}\\
        & s'= \frac{1-\tanh(J_{12}-J_{23}+h_2)}{2}, \quad t'=\frac{1-\tanh(J_{13}+J_{23}+h_3)}{2}, \quad u'=\frac{1-\tanh(J_{13}-J_{23}+h_3)}{2}\\
    \end{aligned}
\end{equation}
Note that $x+x'=1, \ x \in\{p, q, r, s, t, u\}$. To illustrate the procedure, let's examine the construction of following entry $W_{81}$. Using Gibbs sampling, we get
\begin{equation}
    \begin{aligned}
        W_{81} &= P( \{1,1,1\} \longleftarrow  \{-1,-1,-1\})\\
        &= \underbrace{\frac{1+\tanh(J_{12}m_2+J_{13}m_3+h_1)}{2}}_{m_2=-1, m_3=-1} \times \underbrace{\frac{1+\tanh(J_{21}{ m_1}+J_{23}m_3+h_2)}{2}}_{{ m_1=1}, m_3=-1} \times \underbrace{\frac{1+\tanh(J_{31}m_1+J_{32}{m_2}+h_3)}{2}}_{m_1=1, { m_2=1}}\\
        &= p'st
    \end{aligned}
\end{equation}
Note the use of updated value of $m_1$ in updating $m_2$. Similarly, we used the updated value of $m_1$ and $m_2$ in updating $m_3$. This is basically indicating that p-bits are updated sequentially. Following the same procedure, the rest of $W$ entries can be filled. This $W$ matrix is used now to train the AND gate with analytical derivatives.

We illustrate the procedure for training an AND gate, which can be extended in the same manner to a full-adder. The training is accomplished in three main steps:
\begin{enumerate}
    \item[(i)] Construct the transition matrix. We define the probability transition matrix $W$ using the coupling weights $\{J_{12}, J_{23}, J_{13}\}$ and biases $\{h_1,h_2,h_3\}$, as depicted in Supplementary Fig.~\ref{fig:AND_gate}(a). For a network with $p$ layers, one may assign separate $W$ (and hence distinct sets of $J$ and $h$) for each layer.
    
    \item[(ii)] Specify a loss function. We adopt the loss function given by Eq.~\eqref{algo1_eq4}, evaluated over the truth table states in Supplementary Fig.~\ref{fig:AND_gate}(b). For the AND gate, these states are $\mathcal{X} = \{(0,0,0), (0,1,0), (1,0,0), (1,1,1)\}$.

    \item[(iii)] Perform gradient descent. Using gradient-descent we iteratively update each weight and bias until convergence. 
\end{enumerate}

For simplicity, we set $p=1$ (a single layer), which suffices to obtain the desired states. We also choose $\boldsymbol{\rho}_0 = \begin{bmatrix}1 & 0 & \cdots & 0\end{bmatrix}^\top$ as the initial configuration, though in principle any initial state may be used. Notably, the final optimal parameters will be valid only for this specific initial choice, although the approach generalizes to arbitrary $\boldsymbol{\rho}_0$.

Carrying out the above steps leads to
\begin{equation}
\begin{aligned}
\underbrace{
\begin{bmatrix}
    prt & qst & q'rt & p'st & prt & qst & q'rt & p'st \\
    prt' & qst' & q'rt' & p'st' & prt' & qst' & q'rt' & p'st' \\
    pr'u & qs'u & q'r'u & p's'u & pr'u & qs'u & q'r'u & p's'u \\
    pr'u' & qs'u' & q'r'u' & p's'u' & pr'u' & qs'u' & q'r'u' & p's'u' \\
    p's'u' & q'r'u' & qs'u' & pr'u' & p's'u' & q'r'u' & qs'u' & pr'u' \\
    p's'u & q'r'u & qs'u & pr'u & p's'u & q'r'u & qs'u & pr'u \\
    p'st' & q'rt' & qst' & prt' & p'st' & q'rt' & qst' & prt' \\
    p'st & q'rt & qst & prt & p'st & q'rt & qst & prt
\end{bmatrix}
}_{\displaystyle W}
\underbrace{
\begin{bmatrix}
    1\\
    0\\
    0\\
    0\\
    0\\
    0\\
    0\\
    0
\end{bmatrix}
}_{\displaystyle \boldsymbol{\rho}_0}
\;=\;
\underbrace{
\begin{bmatrix}
    prt\\
    prt'\\
    pr'u\\
    pr'u'\\
    p's'u'\\
    p's'u\\
    p'st'\\
    p'st
\end{bmatrix}
}_{\displaystyle \boldsymbol{\rho}_p},
\end{aligned}
\end{equation}
where $\boldsymbol{\rho}_p$ then enters the loss function as follows:
\begin{equation}
\begin{aligned}
\mathcal{L}(\boldsymbol{\theta}) 
&= -\sum_{{\{m\}} \in \mathcal{X}} \ln\bigl[\boldsymbol{\rho}_p({\{m\}};\boldsymbol{\theta})\bigr]\\
&= -\Bigl[\ln\bigl(prt\bigr)+\ln\bigl(pr'u\bigr)+\ln\bigl(ps'u'\bigr)+\ln\bigl(p'st\bigr)\Bigr],
\end{aligned}
\end{equation}
with parameter vector $\boldsymbol{\theta} \;=\;\bigl[J_{12}\;\;J_{23}\;\;J_{13}\;\;h_{1}\;\;h_{2}\;\;h_{3}\bigr]^\top.$

Its derivatives with respect to each parameter are given by:

\begin{subequations}
\begin{equation}
\begin{aligned}
\frac{\partial \mathcal{L}}{\partial J_{12}}
&= \,\text{sech}\bigl(h_{2} + J_{12}\bigr)\,
\Bigl\{
  -\,\text{sech}\bigl(h_{1} + J_{12} + J_{13}\bigr)
  \Bigl[
      \cosh\bigl(h_{1} - h_{2} + J_{13}\bigr)
      + \cosh\bigl(h_{1} + h_{2} + 2\,J_{12} + J_{13}\bigr)\nonumber\\[-6pt]
&\qquad\qquad\quad\ 
      - 4\,\sinh\bigl(h_{1} + h_{2} + 2\,J_{12} + J_{13}\bigr)
  \Bigr]
  \;+\;
  2\,\sinh\bigl(J_{23}\bigr)\,
  \Bigl[
    -\,\text{sech}\bigl(h_{2} + J_{12} - J_{23}\bigr)
    + \text{sech}\bigl(h_{2} + J_{12} + J_{23}\bigr)
  \Bigr]
\Bigr\}
\end{aligned}    
\end{equation}
\vspace{-1.5pc}
\begin{equation}
\begin{aligned}
\frac{\partial \mathcal{L}}{\partial J_{13}}
&=\;
2\,\text{sech}\!\bigl(h_{3}+J_{13}\bigr)\,
\Bigl\{
  \sinh(J_{23})\bigl(\text{sech}[\,h_{3}+J_{13}+J_{23}\,]
    \;-\;\text{sech}[\,h_{3}+J_{13}-J_{23}\,]\bigr)\,
  \nonumber\\
&-\,\text{sech}\!\bigl(h_{1}+J_{12}+J_{13}\bigr)\Bigl(
    \cosh\bigl(h_{1}-h_{3}+J_{12}\bigr)
    \;+\;\cosh\bigl(h_{1}+h_{3}+J_{12}+2\,J_{13}\bigr)
    \nonumber
    -\,2\,\sinh\bigl(h_{1}+h_{3}+J_{12}+2\,J_{13}\bigr)
  \Bigr)
\Bigr\}
\end{aligned}    
\end{equation}
\vspace{-1.5pc}
\begin{equation}
    \begin{aligned}
        \frac{\partial \mathcal{L}}{\partial J_{23}}
&=\;
-2\tanh\bigl(h_{2}+J_{12}-J_{23}\bigr)
  \;-2\tanh\bigl(h_{3}+J_{13}-J_{23}\bigr)
  +2\tanh\bigl(h_{2}+J_{12}+J_{23}\bigr)
  \nonumber +2\tanh\bigl(h_{3}+J_{13}+J_{23}\bigr)
  -2
    \end{aligned}
\end{equation}
\vspace{-1.5pc}
\begin{equation}
    \begin{aligned}
        \frac{\partial \mathcal{L}}{\partial h_{1}}
= 4\tanh\bigl(h_{1}+J_{12}+J_{13}\bigr)
\;-\;2\nonumber
\end{aligned}
\end{equation}
\vspace{-1.5pc}
\begin{equation}
    \begin{aligned}
        \qquad\quad 
        \frac{\partial \mathcal{L}}{\partial h_{2}}
=  2\text{ sech}\bigl(h_2+ J_{12}-J_{23}\bigr) \text{sech}\bigl(h_2+ J_{12}+J_{23}\bigr)\text{sinh}\bigl(2(h_2+ J_{12})\bigr)  \nonumber
\end{aligned}
\end{equation}
\vspace{-1.5pc}
\begin{equation}
    \begin{aligned}
        \frac{\partial \mathcal{L}}{\partial h_{3}}
= -\,\,\text{sech}\bigl(h_{3}+J_{13}-J_{23}\bigr)\,
        \text{sech}\bigl(h_{3}+J_{13}+J_{23}\bigr)\Bigl[
  \cosh\bigl(2\,(h_{3}+J_{13})\bigr)
  \;+\;\cosh\bigl(2\,J_{23}\bigr)
  \;-\;2\,\sinh\bigl(2\,(h_{3}+J_{13})\bigr)
\Bigr]\nonumber
    \end{aligned}
\end{equation}
\end{subequations}

Finally, the gradient-descent update rules are:
\begin{equation}
    \begin{aligned}
    &J^{(t+1)}_{ij} \;\leftarrow\; J^{(t)}_{ij}\;-\;\eta\, \frac{\partial \mathcal{L}}{\partial J^{(t)}_{ij}}, \quad  1\leq i<j\leq 3,   \\
    &h^{(t+1)}_{i} \;\leftarrow\; h^{(t)}_{i}\;-\;\eta\, \frac{\partial \mathcal{L}}{\partial h^{(t)}_{i}}, \quad 1\leq i \leq 3,
    \end{aligned}
\end{equation}
where $\eta$ is a hyperparameter (step size). Each iteration refines the parameters and thus updates the transition matrix $W$, yielding the new distribution $\boldsymbol{\rho}_p^{(t+1)}$. The stopping criterion is set to be either a fixed iteration budget or a certain tolerance in the gradient.

Using the method explained above with $\eta= 0.02$, maximum iterations of $2000$, tolerance in the gradient of $10^{-6}$, and uniform random initialization of weights and biases, specifically, $\boldsymbol{\theta} \sim \text{rand}_{\displaystyle \text{u}}[-0.5, 0.5]$, we get the following rounded optimal weights and biases:

\begin{equation}
    J_{\text{optimal}} = \begin{pmatrix}
    0 & 0 & 2.25 \\
    0 & 0 & 2.25 \\
   2.25 & 2.25 & 0 \\    
    \end{pmatrix}, \quad h_{\text{optimal}} = \begin{bmatrix}
    2.25 \\
    2.25 \\
    -2.25 \\    
    \end{bmatrix}
\end{equation}

The training loss across optimization iterations and the final distribution generated using the optimal parameters are shown in Supplementary Fig.~\ref{fig:AND_gate}. 

\begin{figure}[htp!]
    \centering
    \includegraphics[width=1\linewidth]{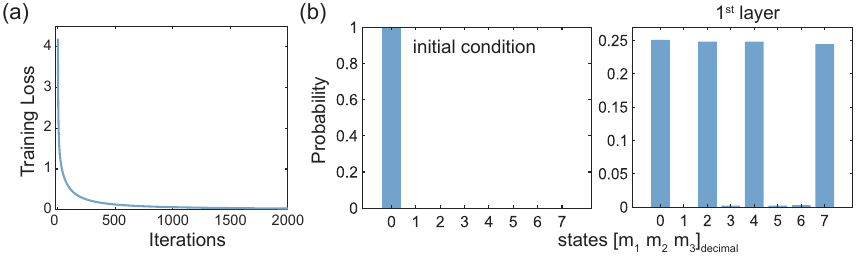}
    \caption{(a) Cost versus optimization iterations for the probabilistic  AND-gate, trained using the negative log-likelihood and gradient descent.  (b) The system PDF evolution using the optimal weights and biases.}
    \label{fig:AND_gate_results}
\end{figure}

\pagebreak

\section{FPGA Implementation of On-chip Annealing}

In the main paper, we present experimental results obtained using a hybrid classical-probabilistic computing system. Here, we provide the technical details of the FPGA-based p-computer implemented on the Xilinx VCU128 data center accelerator card.

\begin{figure}[htp!]
    \centering
    \includegraphics[width=0.85\linewidth]{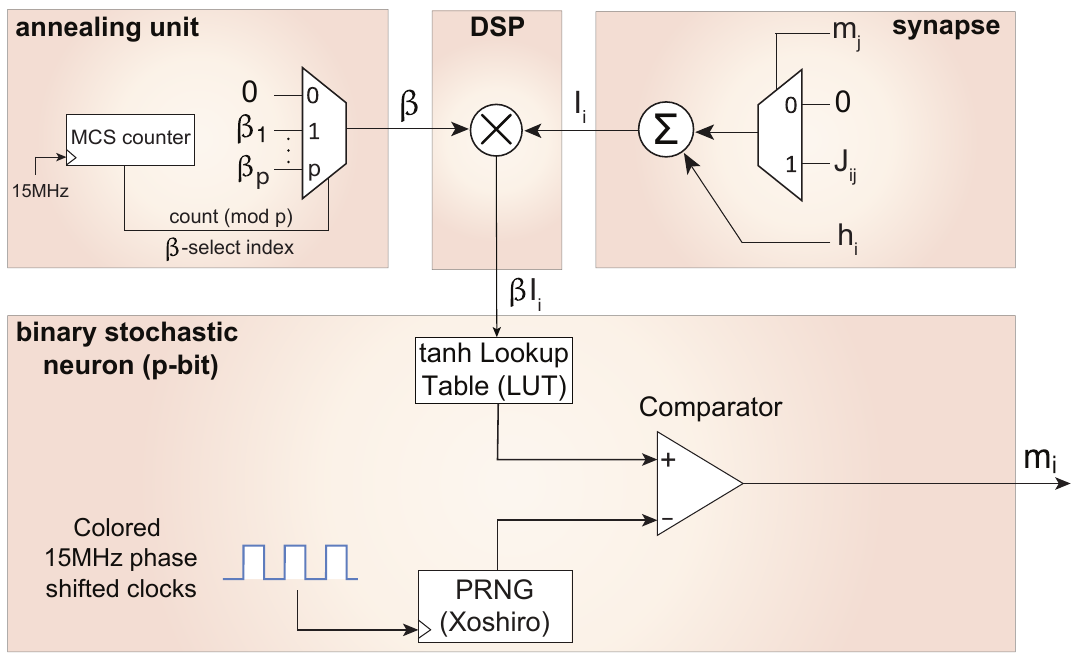}
    \caption{ Probabilistic computer architecture with on-chip annealing. The annealing unit is a $p\times 1$ multiplexer controlled by a counter that updates $\beta$ value after MCSs budget is elapsed. The synapse block implements Eq.~\eqref{algo1_eq2} by using a $2\times 1$ multiplexer and finding the sum over all states $\{m\}$. The DSP slice carries out the multiplication of $\beta$ and the synapse input $I_i$. The BSN unit implements Eq.~\eqref{algo1_eq1}, uses lookup table for $\tanh$, Xoshiro \cite{xoshiro} as the pseudorandom number generator, and a comparator to update the p-bit states. }
    \label{fig:online_annealing}
\end{figure}

The p-computer architecture with on-chip annealing (shown in Supplementary Fig.~\ref{fig:online_annealing}) consists of four main blocks as follows:
\begin{itemize}
    \item Annealing unit: outputs $\beta$ based on the MCS counter. The MCS counter increases once a fixed MCSs budget is elapsed. The value of the counter is then used to select the value of the $\beta$. After exhausting all $\beta$ values for all layers, the counter value will start from the beginning for a new experiment. For the experiments to be independent, the initial $\beta$ value is set to zero to randomize the p-bits. The fixed-point precision used here is $s\{4\}\{5\}$, where $s$ denotes the signed bit, and the values in the square brackets represent the integer and fraction bits, respectively. 
    \item Synapse: outputs $I_i$ based on the states, weights, and biases. For each $m_j$, the multiplexer either chooses $J_{ij}$ or zero. All results are then added at the end along with the bias to calculate the input to node $i$. The fixed-point precision used here for weights and biases is $s\{4\}\{5\}$. 
    \item DSP: outputs the multiplication of $I_i$ and $\beta$. 
    \item Binary stochastic neuron: outputs the updated binary state of node $i$, $m_i$. The output of the DSP block is taken to the LUT to find the corresponding $\tanh$ value. This value is compared with a random number generated via Xoshiro \cite{xoshiro} using a comparator.  
\end{itemize}

\pagebreak

\section{Parameters Trained on Small-size SK Instances for Larger Problems}
\begin{figure}[htp!]
    \centering
    \includegraphics[width=\linewidth]{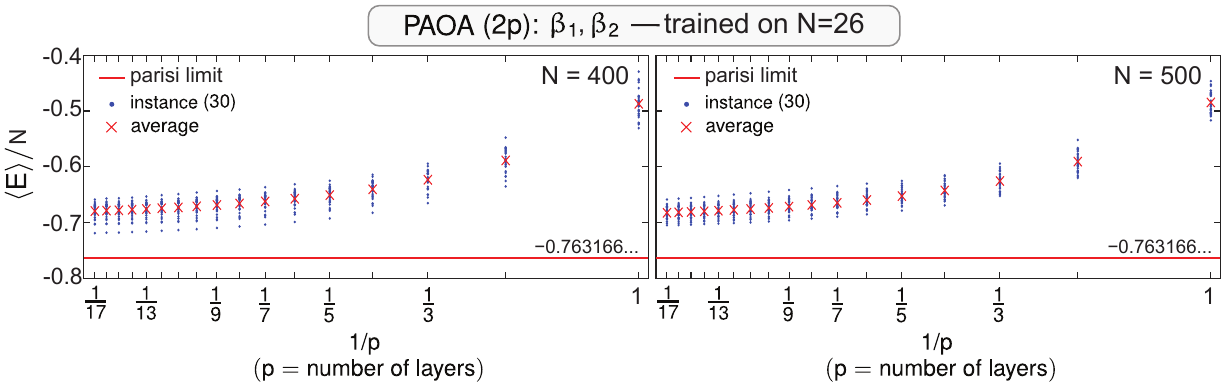}
    \caption{The average energy per spin \(\langle E \rangle/N\) for 30 random large SK instances of size \(N \in \{400, 500\}\) using parameters trained on $N$=$26$ and $10^6$ independent experiments. }
    \label{fig:Large_SK_PAOA}
\end{figure}
In this section, we evaluate the generalization of parameters trained on $N$ = $26$-spin SK instances by applying them to significantly larger problems with $N$ = $400$ and $N$ = $500$, without any additional fine-tuning. Fig.~\ref{fig:Large_SK_PAOA} shows the average energy per spin for these larger Sherrington–Kirkpatrick systems, plotted alongside the Parisi value, which represents the ground state energy in the thermodynamic limit. As the number of PAOA layers increases, the average energy decreases and the instance-to-instance variation narrows and concentrates-more so than in the $N$ = $26$ case. This behavior suggests that the optimized schedules generalize effectively to larger system sizes, and that the remaining fluctuations are primarily due to finite-size effects. It is worth noting that problem sizes of this scale are currently intractable for existing QAOA hardware, underscoring a key practical advantage of PAOA.

\section{Methodology of Schedule Privatization in the L\'evy SK Model}
\begin{figure}[htp!]
    \centering
    \includegraphics[width=0.5\linewidth]{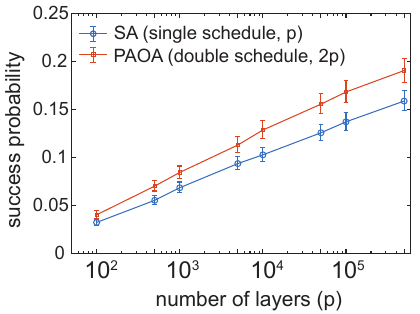}
    \caption{Average success probability over 500 instances comparing single-schedule SA (blue) and double-schedule PAOA (red) for $N$=$50$ and a $50\%$-$50\%$ split. Error bars are $95\%$ confidence intervals from $10^5$ bootstrap samples with replacement.}
    \label{fig:two_schedules_50_50}
\end{figure}

To evaluate the performance difference between conventional single-schedule simulated annealing and a multi-schedule annealing scheme applied to the SK model with Lévy bonds \cite{boettcher2012ground}, the following two-step procedure is adopted:
\begin{enumerate}
    \item Baseline schedule optimization. We perform a grid search over final inverse-temperature values to identify an optimal single geometric annealing schedule at shallow depth ($p$=$17$). The initial inverse temperature (corresponding to high temperature) is held fixed at a small value, while the final value is varied to minimize the average energy per spin across multiple instances and multiple runs.
    
    \item Privatized schedule generation. Using the optimized schedule from step 1, denoted as $\beta_1$, we introduce schedule heterogeneity by applying a multiplicative spacing factor $\Delta$, which is guided by the separation exhibited by PAOA trained double schedules. The original $\beta_1$ is assigned to heavy nodes (i.e., those strongly coupled), while the light nodes receive a schedule scaled by $(1 + \Delta)$, i.e., $\beta_2 = (1 + \Delta)\beta_1$, where $\Delta$ is chosen to reflect the separation that shallow PAOA averaged schedule exhibited when trained on $N$=$50$ with 50 instances in (see Fig.~\ref{fig:SK_Levy_Model}b in the main text). We note that the split -used in Fig.~\ref{fig:SK_Levy_Model} in the main text- into heavy and light groups was done such that top $20\%$ (heaviest 10 nodes) are assigned $\beta_1$, and the rest is assigned $\beta_2$. As shown in Fig.~\ref{fig:two_schedules_50_50}, when using a $50\%$-$50\%$ split, we observe similar behavior shown in Fig.~\ref{fig:SK_Levy_Model}d in the main text.
\end{enumerate}

This procedure follows closely the guidance of PAOA, which suggests scaling the optimized single schedule up and assigning it to the light nodes rather than scaling down the best single annealing schedule and assigning it to the heavy nodes, while keeping the light nodes running at the best annealing schedule. This choice is determined solely by PAOA owing to its learning capability.  

The impact of this scheduling strategy is assessed by comparing the success probability of solving 500 instances, showing that the double schedules learned by PAOA have a consistent and a statistical advantage over a single annealing schedule.

\end{document}